\documentclass[11pt,aps,showpacs,nofootinbib,longbibliography]{revtex4-2}

\pdfoutput=1

\usepackage{comment}
\usepackage[dvipsnames]{xcolor}
\usepackage{amssymb,amsmath,latexsym,mathrsfs}
\usepackage[USenglish,american]{babel}
\usepackage{bm}
\usepackage{amsfonts}
\usepackage{graphicx}
\usepackage{epstopdf}
\usepackage{hyperref}
\usepackage{array}
\usepackage[utf8]{inputenc}
\usepackage{soul}
\usepackage{color}
\usepackage{slashed}
\usepackage{csquotes}
\usepackage[T1]{fontenc}

\usepackage{footnote}

\everymath{\displaystyle}

\newcommand{\beq}{\begin{equation}}
\newcommand{\beqn}{\begin{eqnarray}}
\newcommand{\eeq}{\end{equation}}
\newcommand{\eeqn}{\end{eqnarray}}

\begin{document}

\title{Symmetric multifield oscillons
  }
\author{
Fabio van Dissel$^{1, 2}$ and Evangelos~I.~Sfakianakis$^{1,2,3}$
}
\email{\baselineskip 11pt Email addresses: fvdissel@ifae.es; esfakianakis@ifae.es}
\affiliation{
$^1$ Institut de F\'isica d'Altes Energies (IFAE), The Barcelona Institute of
Science and Technology (BIST), Campus UAB, 08193 Bellaterra, Barcelona, Spain
\\
$^2$Institute Lorentz of Theoretical Physics, Leiden University, 2333 CA Leiden, The Netherlands
\\
$^3$ Nikhef, Science Park 105, 1098 XG Amsterdam, The Netherlands
\\
}
\begin{abstract}
Oscillons are long-lived, spatially localized  field configurations, which are supported by attractive non-linearities in the scalar potential.  
 We study oscillons comprised of multiple interacting fields, each  having an identical potential with quadratic, quartic and sextic terms. We consider  quartic interaction terms of either attractive or repulsive nature. 
In the two-field case, we construct semi-analytical  oscillon profiles for different values of the potential parameters  and coupling strength using  the two-timing small-amplitude formalism.
 We use analytical and numerical techniques to explore the basin of attraction of stable oscillon solutions  and show that, depending on the initial perturbation size, unstable oscillons can either completely disperse or relax to the closest stable configuration. 
 %
 %
 We generalize our analysis to multi-field oscillons and show that the governing equations for their shape and stability can be mapped to the ones arising in the two-field case.
 Finally, we study the emergence of multicomponent oscillons in one and three spatial dimensions, both numerically and through Floquet theory.
 \end{abstract}

\maketitle

\tableofcontents

\section{Introduction}
\label{sec:Intro}

Oscillons are part of a large family of soliton-like structures. 
The stability of these solutions can be traced back to non-linear terms in the theory, since a linear theory will in general force rapid dispersion of any localized wave-packet.  In general, solitons can be classified according two characteristics. They are either completely static (ignoring propagation) or oscillate during their lifetime. Furthermore, their stability is either due to some conserved charge or to interactions between nonlinearities and dispersive effects. 

Among
such non-linear dynamics, oscillons
\cite{Bogolyubsky:1976nx,Bogolyubsky:1976sc,Gleiser:1993pt,Copeland:1995fq,Kasuya:2002zs,Amin:2010jq, Kawasaki:2015vga,Amin:2013ika,Ibe:2019vyo, Sfakianakis:2012bq},
localized long-lived objects, have  attracted significant attention in the early Universe cosmology community.
Oscillons can arise in scalar field models where the potential is quadratic near the minimum and becomes shallower than quadratic (``flattens out'') at larger field values. Currently, inflationary models that are preferred by observations are models that contain plateau-type potentials \cite{Kallosh:2013hoa,Kallosh:2013yoa,Galante:2014ifa,Broy:2015qna}, which satisfy the necessary condition needed to support oscillons.
Furthermore, numerical simulations have shown that the post-inflationary fragmentation of the inflaton in such models leads to copious production of oscillons \cite{Amin:2011hj}.
  
Recently oscillons have seen renewed interest. Numerical simulations have revealed that oscillon formation after inflation is accompanied by the generation of significant amounts of gravitational waves \cite{Zhou:2013tsa,Antusch:2016con,Liu:2017hua,Lozanov:2017hjm,Amin:2018xfe,Kitajima:2018zco,Liu:2018rrt,Lozanov:2019ylm, Hiramatsu:2020obh}. Since gravitational waves can be one of the very few tools able to probe the end of inflation, the dynamics of oscillon formation, evolution and stability are becoming an essential part of early Universe cosmology.

Our current understanding of fundamental physics suggests  that multiple scalar fields are likely to be present at high energies. In spite of this, most of the work on oscillons has focused on single-field models, ignoring decay channels of oscillons to other fields and --more interesting-- the dynamics that can lead to oscillons comprised of multiple fields. Numerical simulations have uncovered two-field oscillons arising after hybrid inflation \cite{Gleiser:2011xj} and in an $SU(2)$ gauged Higgsed model \cite{Farhi:2005rz}, for which a semi-analytical construction of the observed oscillons was given in Ref.~\cite{Sfakianakis:2012bq}, along with a detailed study of their parameter dependent stability. 
The analysis of oscillons in an Abelian Higgs model can be found in Refs.~\cite{Achilleos:2013zpa, Diakonos:2014hra}.
While a type of composite Q-balls have been found in Ref.~\cite{Copeland:2014qra} and further studied in Ref.~\cite{xie2021chargeswapping}, their comparison to composite oscillons is beyond the scope of the present work.
Finally, oscillons were  found in an  $SU(2)\times U(1)$ model \cite{Graham:2006vy, Graham:2007ds}, inspired by the electroweak sector of the Standard Model. In the case of the SU(2) Higgsed oscillon, a particular mass ratio between the Higgs and W fields was required for the oscillon to be stable. In particular, oscillons where the Higgs mass was almost twice the W mass were found to be stable over very long time-scales.  This observation (which was examined in detail in Ref.~\cite{Sfakianakis:2012bq}) can be taken to imply that rather special conditions need to exist for multi-field oscillons to arise.

The current work aims to provide a  detailed look into the conditions for the existence and stability of multi-component oscillons, albeit in a simplified model.
 We organize our presentation as follows. 
In Section \ref{sec:Model} we describe the two-field model that we consider. 
In Section \ref{sec:expansion} we use the small-amplitude two-timing analysis to construct solutions for composite oscillons, comprised of the two fields oscillating in a phase-locked configuration. 
Section \ref{sec:stability} contains the stability analysis of oscillons against long-wavelength perturbations, which is an extension of the Vakhitov-Kolokolov criterion in the case of two interacting fields with identical potential structure, going beyond the specific potential that we chose.
In section \ref{sec:lifetime} we find qualitative and quantitative estimates for the lifetime of stable oscillons that were found in previous sections, following the methods developed  in Refs.~\cite{Zhang_2020classdec, Ibe_2019, Mukaida_2017}.
In section \ref{sec:Nfields} we generalize our analysis to a model containing an arbitrary number of interchangeable interacting fields. 
Finally, Section~\ref{sec:emergence} explores the emergence of multi-field oscillons from a variety of initial conditions.
We provide our conclusions and outlook on future work in Section \ref{sec:Summary}.


\section{Model}
\label{sec:Model}

We consider a model of two identical real scalar fields, each with a quadratic-quartic-sextic potential and a two-to-two interaction term, which can be either  attractive or repulsive. This form of the potential was first analyzed for single-field oscillons in 
 Ref.~\cite{Amin:2010jq}, where a large sextic term was introduced to stabilize three-dimensional oscillons in a symmetric quadratic-quartic potential. We ignore the expansion of the Universe\footnote{We will present numerical results for an expanding Universe in one spatial dimension in Section~\ref{sec:emergence}.} and restrict ourselves to an action consisting of the two scalar fields on a Minkowski background with metric signature $(-,+,+,+)$:
\beq
S = -\int d^3x dt \left [
\sum_{I=1,2} \left ( {1\over 2} \partial_\mu \phi^I\partial^\mu \phi^I + {1\over 2} m^2 \left (\phi^I \right )^2
- {\lambda\over 4} \left  (\phi^I \right )^4 + \frac{g}{6}\left (\phi^I \right  )^6
\right ) -\frac{\Lambda}{2}(\phi^1)^2 (\phi^2)^2
\right ]
\, .
\label{eq:Action}
\eeq

Before proceeding, it is worth noting both the restrictions that the action of Eq.~\eqref{eq:Action} poses, as well as some theoretical motivation for considering it. We  draw our example from the well-motivated area of $\alpha$-attractors, specifically the T-model potential. Since metrics on hyperbolic spaces can be formulated in many different forms, following M\"obius transformations \cite{Carrasco:2015uma}, we follow the metric definition used in Refs.~\cite{Krajewski:2018moi, Iarygina:2018kee, Iarygina:2020dwe}
\beq
ds^2 = d\chi^2 + e^{2b(\chi)} d\phi^2 \, ,
\eeq
where $b(\chi) = \log \left (\cosh(\beta\chi) \right )$. The two-field potential for the T-model in this field basis is
\beq
V(\phi,\chi) = \alpha\mu^2 \left (
{
\cosh(\beta\phi ) \cosh(\beta\chi )-1 \over
\cosh(\beta\phi ) \cosh(\beta\chi )+1
}
\right ) \left (
\cosh(\beta\chi)
\right)^{2/\beta^2}
\, ,
\eeq
where $\beta =\sqrt{2/3\alpha}$.

For small value of $\alpha$, the potential around the origin is expanded as
\beqn
\nonumber
V(\phi,\chi) &\simeq &
{\mu^2 \over 6}\chi^2 - {\mu^2 \over 54\alpha} \chi^4 + {17 \mu^2 \over 9720 \alpha^2}\chi^6
+ {\mu^2 \over 6}\phi^2 - {\mu^2 \over 54\alpha} \phi^4 + {17 \mu^2 \over 9720 \alpha^2}\phi^6
\\&&+ {\mu^2 \over 6}\phi^2 \chi^2  - {\mu^2 \over 16 \alpha^2} \phi^2 \chi^2 (\phi^2+\chi^2) 
+ ...
\eeqn
where we included terms up to sextic order in the fields and neglected higher order contributions in $\alpha$ in each term. 
For that we assume that the parameter $\alpha$, which is inversely related to the field-space curvature, is small.
We see that this expansion has the characteristics that we consider in the action of Eq.~\eqref{eq:Action}: two scalar fields with identical potential parameters, a negative quartic term and a very large sextic term (for $\alpha \ll 1$), as well as a quartic coupling. Ref.~\cite{Krajewski:2018moi} reported the formation of oscillons during two-field preheating after inflation on a T-model potential, but did not provide a detailed analysis of their properties.
\begin{figure}
\centering
 \includegraphics[width=\textwidth]{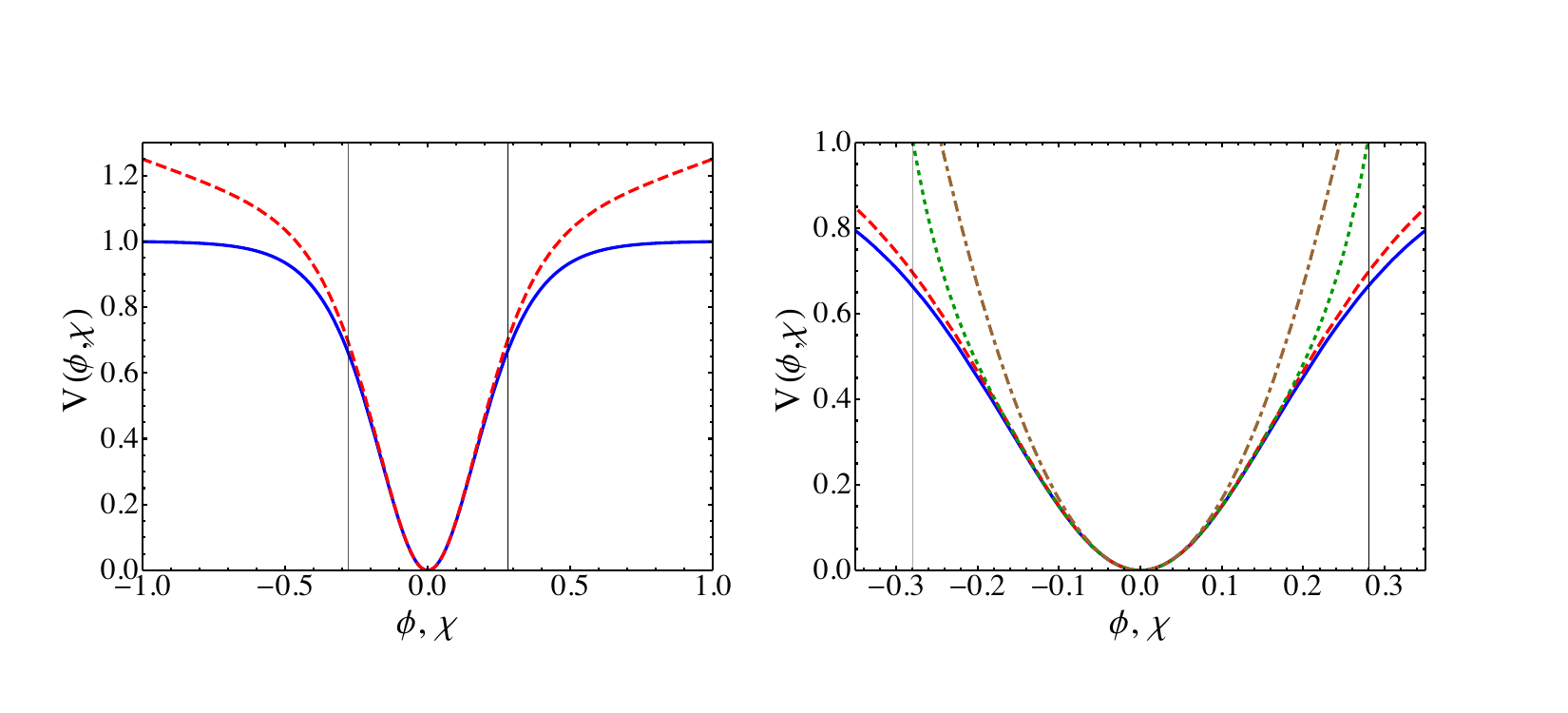} 
\caption{
The T-model potential for $\chi=0$ (blue solid) and $\phi=0$ (red dashed), along with the sextic Taylor expansion (green dotted). The brown dot-dashed curve shows the quadratic term, which is steeper than the total potential, allowing in principle for oscillon formation in both $\phi$ and $\chi$ fields.
The vertical lines show the field value of $\phi$ at the end of inflation. 
}
 \label{fig:Tmodel}
\end{figure}
Of course, the true T-model expansion differs from the idealized action of Eq.~\eqref{eq:Action}, in that it contains sextic interaction terms and it provides less freedom for choosing the various potential coupling strengths (see Fig.~\ref{fig:Tmodel}). Furthermore, the analysis of the T-model must also take into account the non-canonical kinetic structure, leading to the derivative couplings
\beq
{\cal L}_{\rm kin} \supset {\chi^2\over 3\alpha} (\partial \phi)^2 + {2 \chi^4\over 27\alpha^2} (\partial \phi)^2 + ...
\eeq
Oscillons in systems with non-canonical kinetic terms have been considered in the litetarure for single-field systems \cite{Amin:2013ika}. 
 In the present work, we instead focus on the symmetric sextic potential of Eq.~\eqref{eq:Action} and consider canonical kinetic terms for the two fields. An extension of the current analysis  to realistic $\alpha$-attractor models, taking into account the intricacies of derivative couplings, can reveal interesting phenomenology for this family of well-motivated and observationally relevant models and  is left for future work.


\section{Two timing analysis}
\label{sec:expansion}

In this section we construct  approximate profiles of two-field oscillons and compare them to the single-field oscillons  found in Ref.~\cite{Amin:2010jq}.
Eq.~\eqref{eq:Action} contains several  physical parameters, such as the mass of the two fields $m$, the coupling $\Lambda$ and self couplings $g$ and $\lambda$. 
 It is convenient to work with dimensionless space-time variables and fields. This is done by the rescalings $x^\mu \rightarrow\tilde{x^\mu}= x^\mu m$, $\phi^I \rightarrow \tilde{\phi}^I =  m^{-1} \lambda^{1/2} \phi^I$, $g \rightarrow \tilde{g}=(m/\lambda)^2 g$ and $\Lambda \rightarrow \tilde{\Lambda} = (1/ \lambda) \Lambda$. The  Lagrangian is then re-written as 
\beq
\mathcal{L} =\frac{-\lambda}{m^4}\left( \sum_{I=1,2} \left( {1\over 2} \partial_\mu \tilde{\phi}^I\partial^\mu \tilde{\phi}^I + {1\over 2} \left ( \tilde{\phi}^I \right )^2
- {1\over 4} \left  (\tilde{\phi}^I \right )^4 + \frac{\tilde{g}}{6}\left (\tilde{\phi}^I \right  )^6
\right ) -\frac{\tilde{\Lambda}}{2}(\tilde{\phi}^1)^2 (\tilde{\phi}^2)^2\right)
\label{eq:lagdim}
\eeq
For simplicity, in what follows we define $\{ \tilde{\phi}^1,\tilde{\phi}^2 \} \equiv \{ \tilde{\phi}, \tilde{\chi}\}$  and drop all  tildes. 
The Lagrangian of Eq.~\eqref{eq:lagdim} leads to a system of two  coupled equations of motion for $\phi$ and $\chi$.
In this work we focus our attention on spherically symmetric solutions $\phi(\vec x ,t) \to \phi(r ,t)$, leading to the equations of motion
\begin{equation}
\begin{split}
\partial_t^2 \phi - \left (\partial_r^2 + \frac{2}{r}\partial_r \right )\phi + \phi = \phi^3 - g \phi^5 + \Lambda \phi \chi^2
\, ,
\\
\partial_t^2 \chi -\left (\partial_r^2 + \frac{2}{r}\partial_r \right )\chi + \chi = \chi^3 - g \chi^5 + \Lambda \chi \phi^2
\, .
\end{split}
\label{eq:motion}
\end{equation}
In $d$ spatial dimension the above equations are altered by the substitution\footnote{Additionally, in $d$ dimensions the rescalings of the fields will also be different.} $\frac{2}{r}\partial_r  \to\frac{d-1}{r}\partial_r $.
The attentive reader will immediately notice that these equations are symmetrical under exchange of the fields, meaning that sending $\chi \leftrightarrow \pm \phi$ in either of the equations gives back the other. This greatly simplifies the analytical search for oscillons which is more difficult (if at all possible) in  general multi-component systems.

It is well established in the literature that oscillons live on long time- and large spatial scales \cite{Fodor:2009kf, Gleiser:2009ys, Salmi:2012ta}. This suggests a perturbative approach to extract oscillon solutions from the equations of motion, known as the small-amplitude two-timing analysis. The idea is that oscillons exhibit behaviour on two time scales, one capturing the natural frequency of the free field theory and one capturing the correction to this frequency characteristic for the non-linear potential. 
Furthermore, we attempt to capture oscillons which are slowly varying in space, hence are ``broad''.
This behaviour is found by introducing new time and space variables
\beq
\tau = \alpha \epsilon^2 t \, ,~ \rho = \epsilon r
\label{eq:slowtimevar}
\eeq
where $\alpha$ is a free parameter and $\epsilon \ll 1$. 
While we expect all spatial variation to occur on the scale of $\rho$, the time variation occurs on two scales: $t$ and $\tau$. 
The (double) time derivatives in Eqs.~\eqref{eq:motion} must now be interpreted as full time derivatives, leading to the equations
\begin{equation}
\begin{split}
\partial_t^2\phi + 2 \alpha \epsilon^2 \partial_t \partial_{\tau} \phi - \epsilon^2\left (\partial_{\rho}^2 + \frac{2}{\rho}\partial_{\rho}\right )\phi + \phi = \phi^3 - g \phi^5 + \Lambda \phi \chi^2 + O\left (\epsilon^4\right) 
\, ,
\\
\partial_t^2\chi + 2 \alpha \epsilon^2 \partial_t \partial_{\tau}\chi - \epsilon^2 \left (\partial_{\rho}^2 + \frac{2}{\rho}\partial_{\rho} \right )\chi + \chi = \chi^3 - g \chi^5 + \Lambda \chi \phi^2 + O\left (\epsilon^4\right)
\, .
\end{split}
\label{eq:motioncor}
\end{equation}
Lastly, we assume that the amplitude of the oscillon profile  also scales with the expansion parameter $\epsilon$, allowing us to compute the effects of the non-linearities order-by-order.
We therefore  consider solutions of the form
\begin{equation}
\begin{split}
    \phi(x, t) = \sum_{n = 1}\epsilon^n \phi_{n}(\rho, t, \tau) \, ,
\\
    \chi(x, t) = \sum_{n = 1}\epsilon^n \chi_{n}(\rho, t, \tau) \, .
    \end{split}
\label{eq:exp}
\end{equation}
Inserting Eqs.~\eqref{eq:exp} into Eqs.~\eqref{eq:motioncor}, the lowest order equations in $\epsilon$ are
\begin{equation}
\begin{split}
    \partial_t^2 \phi_1 + \phi_1 =& 0 \, ,
    \\
    \partial_t^2 \chi_1 + \chi_1 =& 0 \, ,
\end{split}
\label{eq:first}
\end{equation}
which are identical to the harmonic oscillator equation and capture the main oscillatory behaviour of the oscillon. The equations in the next non-trivial order in $\epsilon$ are
\begin{equation}
\begin{split}
    \partial_t^2 \phi_3 + \phi_3 =& \left(\partial_{\rho}^2 + \frac{2}{\rho}\partial_{\rho} \right )\phi_1 - 2 \alpha \partial_t \partial_{\tau}\phi_1 + \phi_1^3 -  \phi_1^5 + \Lambda \phi_1 \chi_1^2
    \, ,
    \\
    \partial_t^2 \chi_3 + \chi_3 =& \left(\partial_{\rho}^2 + \frac{2}{\rho}\partial_{\rho}\right )\chi_1 - 2 \alpha \partial_t \partial_{\tau}\chi_1 + \chi_1^3 -  \chi_1^5 + \Lambda \chi_1 \phi_1^2
    \, .
\end{split}
\label{eq:third}
\end{equation}
Notice that we used the fact that $g$ is large and have therefore written $g = {1}/{\epsilon^2}$. 
The solutions to the harmonic oscillator Eqs.~\eqref{eq:first} are trivial. Therefore we can write $\phi_1 = \operatorname{Re} \{ A(\rho, \tau)e^{-it} \}$ and $\chi_1 = \operatorname{Re}\{B(\rho, \tau)e^{-it}\}$; where $A(\rho, \tau)$ and $B(\rho, \tau)$ are complex functions\footnote{Note that this procedure is equivalent to going to the non-relativistic limit, where a real scalar field can be written as the real part of a complex wavefunction $\Psi(x, t) \exp^{-imt}$.}. Inserting this solution into Eqs.~\eqref{eq:third} and eliminating secular terms gives us the envelope equations of the system of oscillons
\begin{equation}
\begin{split}
2i\alpha \partial_{\tau}A + \left(\partial_{\rho}^2 + \frac{2}{\rho}\partial_{\rho}\right)A + \frac{3}{4}|A|^2 A + \frac{\Lambda}{2}|B|^2 A +\frac{\Lambda}{4}A^*B^2 -\frac{5}{8}|A|^4 A = 0
\, ,
\\
2i\alpha \partial_{\tau}B + \left(\partial_{\rho}^2 + \frac{2}{\rho}\partial_{\rho}\right)B + \frac{3}{4}|B|^2 B + \frac{\Lambda}{2}|A|^2 B +\frac{\Lambda}{4}B^*A^2 -\frac{5}{8}|B|^4 B = 0
\, .
\end{split}
\label{eq:NLS}
\end{equation}
These equations are of the Nonlinear Schr\"odinger type \cite{Zakharov:1974zf}. They control the behaviour of the oscillon on long time- and large spatial scales. To find solutions of these equations that are localized in space we insert the ``oscillon ansatz''. Since by assumption the oscillon is just some localized structure oscillating in time we should look for solutions of the form, $A(\rho, \tau) = a(\rho) e^{ic_1\tau}$ and $B(\rho, \tau) = b(\rho) e^{ic_2\tau}$; where $a$ and $b$ are real functions determining the spatial character of the oscillon. Due to the symmetry of the potential we can also set $c_1 = c_2 = c$. Inserting this into Eqs. \ref{eq:NLS}, we obtain the profile equations of the oscillons.
\begin{equation}
\begin{split}
- \alpha \, a + \left(\partial_{\rho}^2 + \frac{2}{\rho}\partial_{\rho} \right )a + \frac{3}{4}a^3 + \frac{3\Lambda}{4}b^2a - \frac{5}{8} a^5 = 0 \, ,
\\
- \alpha \, b + \left(\partial_{\rho}^2 + \frac{2}{\rho}\partial_{\rho}\right )b + \frac{3}{4}b^3 + \frac{3\Lambda}{4}a^2b - \frac{5}{8} b^5 = 0 \, ,
\end{split}
\label{eq:prof}
\end{equation}
where we set $c = {1}/{2}$. A few remarks about the parameter $c$ are now in order. $c$ is in essence a free parameter, as long as it is not too large. This is because it can always be absorbed into a redefinition of $\epsilon$. We do require it to be positive, since the oscillon should behave as $a, b \rightarrow 0$ for $\rho\to \infty$, since they are localized solutions. We can therefore set $c = {1}/{2}$ without loss of generality.

An interesting property arising due to the symmetry of the system now becomes apparent. Namely, any localized solution $a(\rho)$ of the equation
\beq
- \alpha \, a + \left(\partial_{\rho}^2 + \frac{2}{\rho}\partial_{\rho}\right )a + \frac{3}{4}(1 + \Lambda)a^3 -  \frac{5}{8} a^5 = 0
\label{eq:profs}
\eeq
directly solves the system of Eqs.~\eqref{eq:prof} if we set $a = b$. We arrive at the same equation if we instead choose $a = - b$, meaning that the two fields oscillate out of phase. It is worth noting that in the one other case in the literature, where two-field oscillons were constructed using the two-timing analysis, Ref.~\cite{ Sfakianakis:2012bq},  the masses of the two fields were chosen to have a $2:1$  ratio, meaning that for half the period of the light field, they were in phase, and for the other half they were out of phase. 

The difficulty of finding oscillon solutions in the coupled system of Eqs.~\eqref{eq:prof} is therefore greatly reduced and can be related to the solutions for single-field oscillons in a quartic-sextic potential, which were studied in Ref.~\cite{Amin:2010jq}.

\subsection{Oscillon profiles}

In order to  solve Eq.~\eqref{eq:profs}, we must first define the boundary conditions. At large distance from the origin, far away from the oscillon core, the non-linear terms become subdominant, hence Eq.~\eqref{eq:profs} can be approximated as $\partial_\rho^2\, a + (2/\rho) \partial_\rho a \approx \alpha a$, leading to $a(\rho) \propto e^{- \sqrt\alpha \rho} / \rho$.
Finally, we require regularity at the origin. Eq.~\eqref{eq:profs} can be re-written as
\beq
{dE_\rho\over d\rho} = -{2\over \rho} \left (
{\partial a\over \partial\rho}
\right)^2 \, ,
\label{eq:dEdrho}
\eeq
where  
\beq
E_\rho = {1\over 2} \left (  {\partial a\over \partial\rho}\right)^2 -{1\over 2} \alpha \, a^2 + {3\over 16} (1+\Lambda) a^4 - {5\over 48}  a^6
\label{eq:energyloc}
\eeq
would be the conserved energy  for a one-dimensional system with the same potential.  The right hand side of Eq.~\eqref{eq:dEdrho} is non-positive, hence the energy of any solution $E_\rho$ will decrease as $\rho\to \infty$.
Since the energy of a localized solution must be zero in the  far-distance regime (this can be seen by inspecting Eq.~\eqref{eq:energyloc}) we conclude that $E_\rho \ge0$ at $\rho\to 0$.
It is best to think about this in terms of the trajectories of solutions in phase space $(a,\partial_\rho a)$. Since we're interested in functions that are smooth at the origin, all trajectories start on the $\partial_\rho a = 0 $ axis. Any trajectory will continuously intersect curves with $E_\rho={\rm const}$, and the constant value of these curves must decrease as $\rho\to \infty$ . The $E_\rho=0$ curve then defines the boundary between localized and non-localized solutions. Namely, a localized solution must intersect this curve in the origin. The requirement that the oscillon is smooth at the origin, meaning $\partial_\rho a|_{\rho=0}=0$, and the fact that $E_{\rho} \geq 0$ there, leads to $\alpha <\alpha_c \equiv (27/160 )(1 + \Lambda)^2$, generalizing the constraint found in Ref.~\cite{Amin:2010jq} for single-field oscillons\footnote{We must note that our definition of the parameter $\alpha$ is related to the square of the parameter $\alpha$  used in Ref.~\cite{Amin:2010jq}.
}, or equivalently  $\Lambda=0$. 
 For $\alpha \to \alpha_c$ the oscillon becomes infinitely wide with an amplitude $a^2(\rho=0) \to {({9}/{10})(1+\Lambda)}$.

\begin{figure}[h] 
		\includegraphics[width=\textwidth]{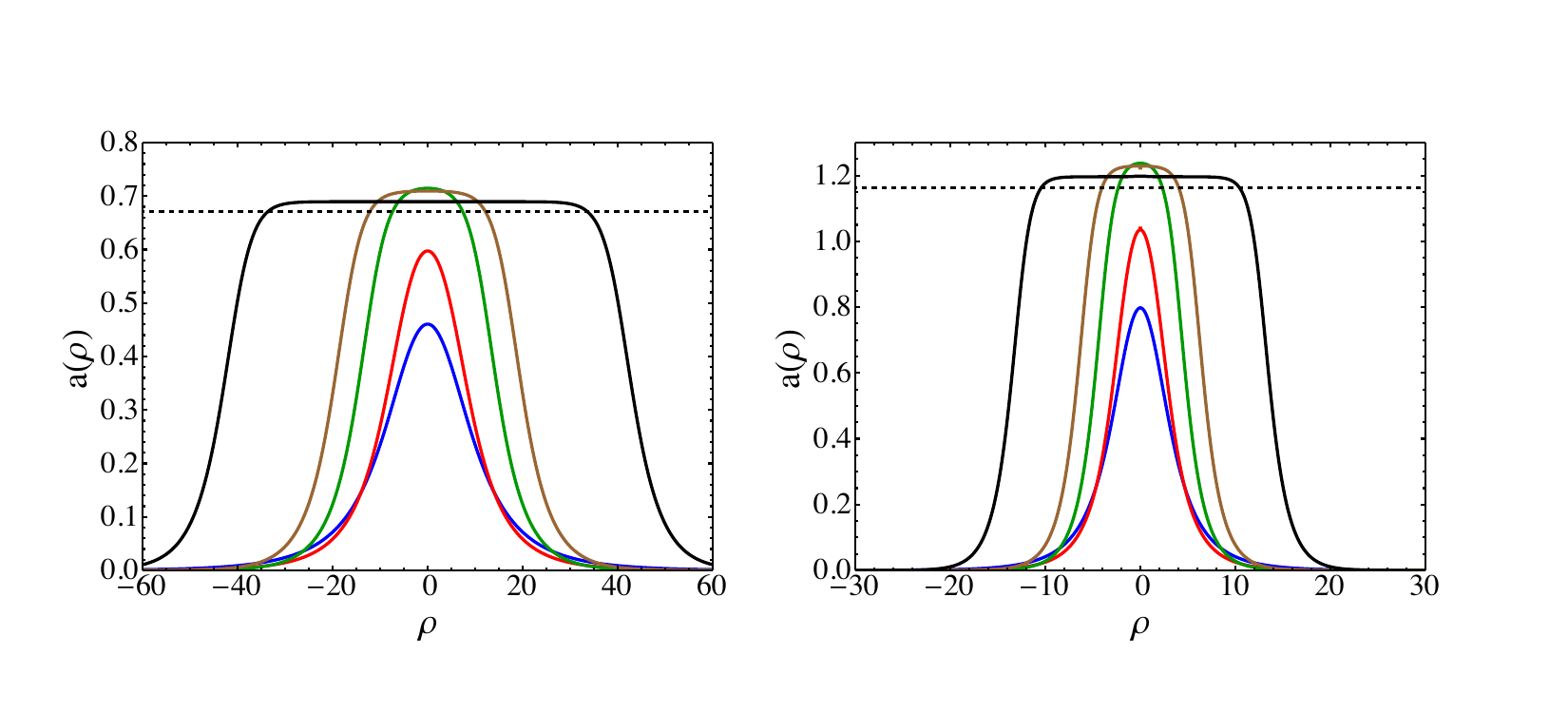} 
		\caption{
		{\it Left}: Oscillon profiles as a function of the rescaled radius $\rho$ for repulsive fields with  $\Lambda=-0.5$  and $\alpha=0.005, 0.01, 0.025, 0.03, 0.037$ (blue, red, green, brown and black, respectively).
			{\it Right}:	Oscillon profiles for attractive fields with $\Lambda=0.5$ 
		and $\alpha=0.045, 0.09, 0.22, 0.27, 0.33$ (blue, red, green, brown and black, respectively).
We see the emergence of a flat-top shape in both cases. Furthermore, we compute that $\alpha_c=0.042$ for  $\Lambda=-0.5$ and $\alpha_c=0.38$ for  $\Lambda=0.5$, leading to,
$a(\rho=0) \simeq 0.67$ and $a(\rho=0) \simeq 1.16$ respectively, meaning that  oscillons can acquire larger amplitudes in a system with  attractive interactions.		} 
\label{fig:profiles} 
\end{figure}

 It turns out that only a countable set $a_n(\rho)$ of localized trajectories can be drawn in phase
space \cite{selffocusing}. Here $n$ is the amount of nodes of the solution. We conclude that there is exactly one
zero-node solution of the profile equation, and thus one zero-mode
oscillon for every choice of $\epsilon$ and $\alpha$. 
We do not pursue solutions with $n\ge1$, since they will have a higher energy than their zero-node counterpart, and thus are expected to be unstable.
We used both  the shooting and relaxation methods in order to find the corresponding oscillon profiles and check our results. Fig.~\ref{fig:profiles} shows some characteristic profiles for $\Lambda=\pm 0.5$. 
For the one-dimensional case, Eq.~\ref{eq:profs} can be solved analytically to give 
\beq
\phi(r) = \sqrt{{9\over 10}(1+\Lambda)}  \sqrt{1-u^2\over 1+ u \cosh(2\sqrt{\alpha} r)} \, ,
\eeq
where $u^2 = 1-{\alpha/\alpha_c} $. For $\Lambda=0$ we recover the results of Ref.~\cite{Amin:2010jq}.

\begin{figure}[h] 
\centering
		\includegraphics[width=.45\textwidth]{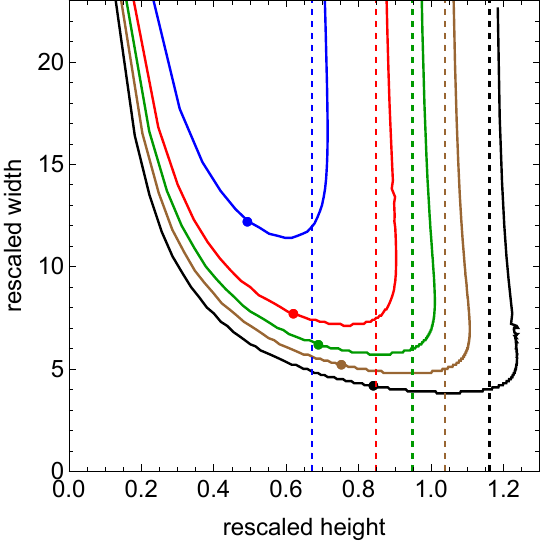}\hspace{.05\textwidth}
			\includegraphics[width=.45\textwidth]{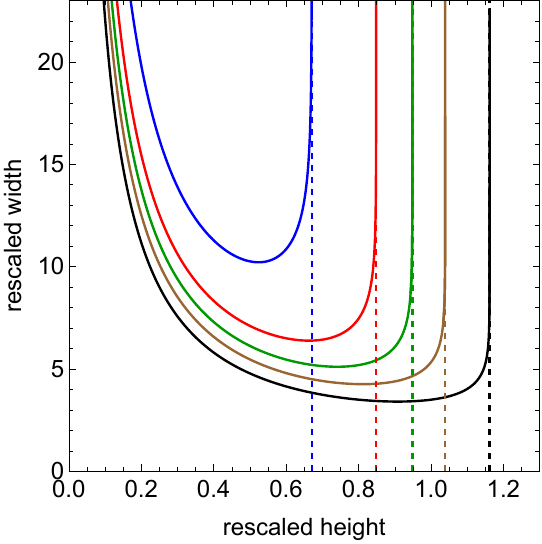} 
		\caption{
		The height-width relation for the oscillon  for $\Lambda=-0.5, -0.2, 0, 0.2, 0.5$ (blue, red, green, brown and black	respectively) in 3 and 1 spatial dimensions (left and right respectively). We see that fields which interact repulsively lead to a smaller range of oscillon amplitudes, leading to no oscillons for $\Lambda\to-1$. On the contrary, attractive interactions allow for taller, as well as wider oscillons. 
		The colored dots  show the point where the oscillons change their stability properties in 3D, as explained in  Section~\ref{sec:stability}.
}\label{fig:heightwidth} 
\end{figure}

Fig.~\ref{fig:heightwidth} shows the  width and height of the oscillons as a function for various interaction strengths between the two fields, attractive or repulsive, rescaled by the small-amplitude parameter $\epsilon$. We see that in the non-interacting case of $\Lambda\to 0$, we recover the single-field results of Ref.~\cite{Amin:2010jq}. 
Furthermore, fields which interact repulsively lead to a smaller range of oscillon amplitudes, ending in no oscillons for $\Lambda\to-1$, as is also evident by the form of $\alpha_c \sim (1 + \Lambda)^2$. On the contrary, attractive interactions allow for  oscillons with a larger amplitude.

\begin{figure}[h] 
		\includegraphics[width=1\textwidth]{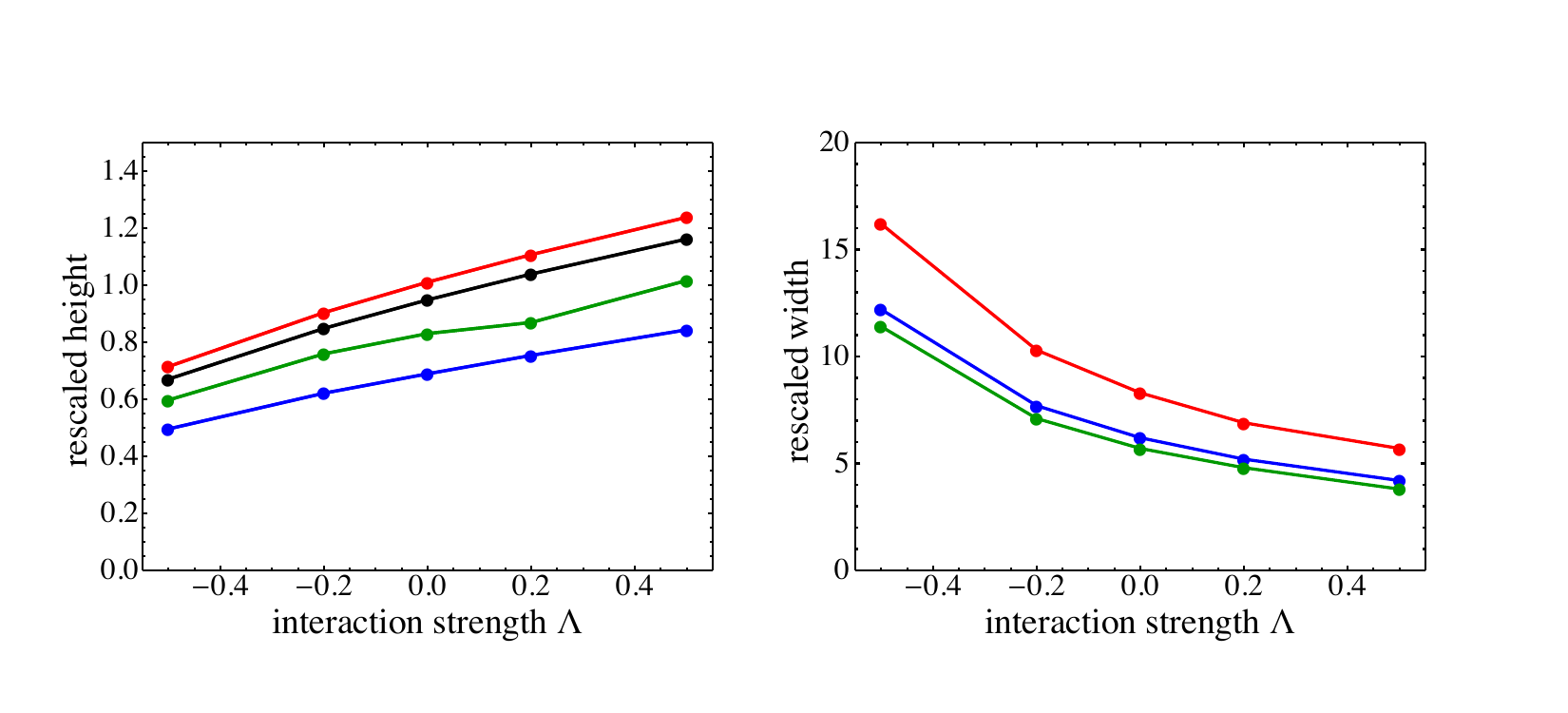} 
		\caption{
		The height and width of 3-dimensional oscillons for the point of marginal stability (blue), the point corresponding to the minimum oscillon width (green) and the point corresponding to the maximum oscillon height (red). The black curve corresponds to the rescaled asymptotic oscillon height $ \sqrt{\frac{9}{10}(1+\Lambda)}$.
}\label{fig:heightswidths} 
\end{figure}

Fig.~\ref{fig:heightswidths} shows several characteristic values for the height and width of oscillons as a function of the coupling strength $\Lambda$. We see that the spread of oscillon heights, calculated as the difference between the minimum height for stable oscillons and the maximum height, grows with $\Lambda$, being almost double for $\Lambda=0.5$ compared to $\Lambda=-0.5$. The spread of the width is infinite, since arbitrarily wide oscillons can exist in theory. For larger values of the interaction strength $\Lambda$, narrower oscillons exist. In particular, the minimum width for oscillons with $\Lambda=-0.5$ is almost triple the corresponding value for $\Lambda=0.5$.

\section{Stability Analysis}
\label{sec:stability}
It has been shown repeatedly in the literature that no true breather solutions like the oscillon can exist in nonlinear systems (with the notable exception of breathers in the one-dimensional Sine-Gordon model).
 In general there will always be (classical) outgoing radiation in the tails of the oscillon. Classically, this radiation is exponentially suppressed, which is why oscillons can be extremely long-lived \cite{Olle:2020qqy}. 
 Quantum mechanical radiation might  play a more important role in real physical systems \cite{Hertzberg:2010yz}, but is beyond the scope of our present work. 
Radiation will in general perturb the oscillon system and it is therefore necessary to assess the stability of oscillons with respect to small perturbations. In the model we are investigating, let us consider small fluctuations  $\delta(x, t), \Delta(x, t) \ll O(\epsilon)$ added to the oscillon solutions as
\beq
\begin{split}
\phi(x, t) &= \phi_{osc}(x, t) + \delta(x,t)
\, , \\
\chi(x, t) &= \chi_{osc}(x, t) + \Delta(x, t)
\, .
\end{split}
\eeq
Plugging these into Eqs.~\eqref{eq:motion}, the equations of motion for the two fields, and keeping only  terms linear in $\delta$ and $\Delta$ leads to\footnote{The terms that were ignored here (the terms only involving $\phi_{osc}$ and $\chi_{osc}$) can by nature of the two-timing analysis not source instabilities. However, in Section~\ref{sec:lifetime} we will show that these terms are a source of outgoing radiation.}
\beq
\begin{split}
\partial_t^2\delta + \delta - (\partial_r^2 + \frac{2}{r}\partial_r)\delta - 3\phi_{osc}^2 \delta + 5g\phi_{osc}^4 \delta - \Lambda \chi_{osc}^2 \delta - 2\Lambda\phi_{osc}\chi_{osc}\Delta=& 0 
\, ,
\\
\partial_t^2\Delta + \Delta - (\partial_r^2 + \frac{2}{r}\partial_r)\Delta - 3\chi_{osc}^2 \Delta + 5g\chi_{osc}^4 \Delta - \Lambda \phi_{osc}^2 \Delta - 2\Lambda\chi_{osc}\phi_{osc}\delta=& 0 
\, .
\end{split}
\label{eq:linearper}
\eeq
 Since the exact shape of the initial perturbation is in essence unpredictable, a full linear stability analysis would require solving Eq.~\eqref{eq:linearper} for arbitrary initial conditions. This requires solving the full Floquet matrix for the coupled nonlinear system. It turns out that, for this system, there is a useful simplification if we only consider perturbations that are about the same size as the oscillon itself. This is in essence an extension of the Vakhitov-Kolokolov criterion that has been used to assess stability of single-component oscillons \cite{Amin:2010jq}.

\subsection{Extension of the V-K criterion}
\label{sec:VK}
In  Section~\ref{sec:expansion} we have already shown that this system supports oscillons of the form 
\beq
\begin{split}
\phi_{osc}(x, t) &=\epsilon \,  \operatorname{Re} \left \{{a(\epsilon x) e^{it(1-\frac{\epsilon^2}{2})}} \right \}=\epsilon \, a(\epsilon x) \cos{\omega t} \, ,
\\
\chi_{osc}(x, t) &=\epsilon  \, \operatorname{Re} \left \{ {b(\epsilon x) e^{it(1-\frac{\epsilon^2}{2})}} \right \}=\epsilon \,  b(\epsilon x) \cos{\omega t} \, ,
\end{split}
\label{eq:phichiosc}
\eeq
where the frequency\footnote{In several oscillon studies, the frequency is taken to be of the form $\omega =\sqrt{1-c\cdot h^2}$, where $h$ is the oscillon height and $c$ is some ${\cal O}(1)$ constant. This matches our expression to lowest order in perturbation theory when $h\propto \epsilon \ll 1$, which is true in the context of the small amplitude analysis used to construct oscillon solutions.} is $\omega = 1 - \alpha {\epsilon^2}/{2}$. The symmetry of the system dictates that we have $a(\rho) =\pm b(\rho)$, as we have discussed in Section~\ref{sec:expansion}. This allows us to turn Eqs.~\eqref{eq:phichiosc} into a single-variable system in order to derive and extend the V-K criterion, similarly to what was done in Ref.~\cite{Amin:2010jq} for single-field oscillons. 
Within the validity of the VK criterion, oscillons are stable with respect to long-wavelength perturbations if and only if $dN/d\alpha > 0$, where
\beq
N = \int a^2(\rho) d^3\rho \, .
\label{eq:Ndef}
\eeq
We present the details of the derivation in Appendix~\ref{app:VK}. Fig.~\ref{fig:VKrelations} shows the regions of stability, based on Eq.~\eqref{eq:Ndef}.

\begin{figure}[ht!]
    \centering
    \includegraphics[width=0.8\textwidth]{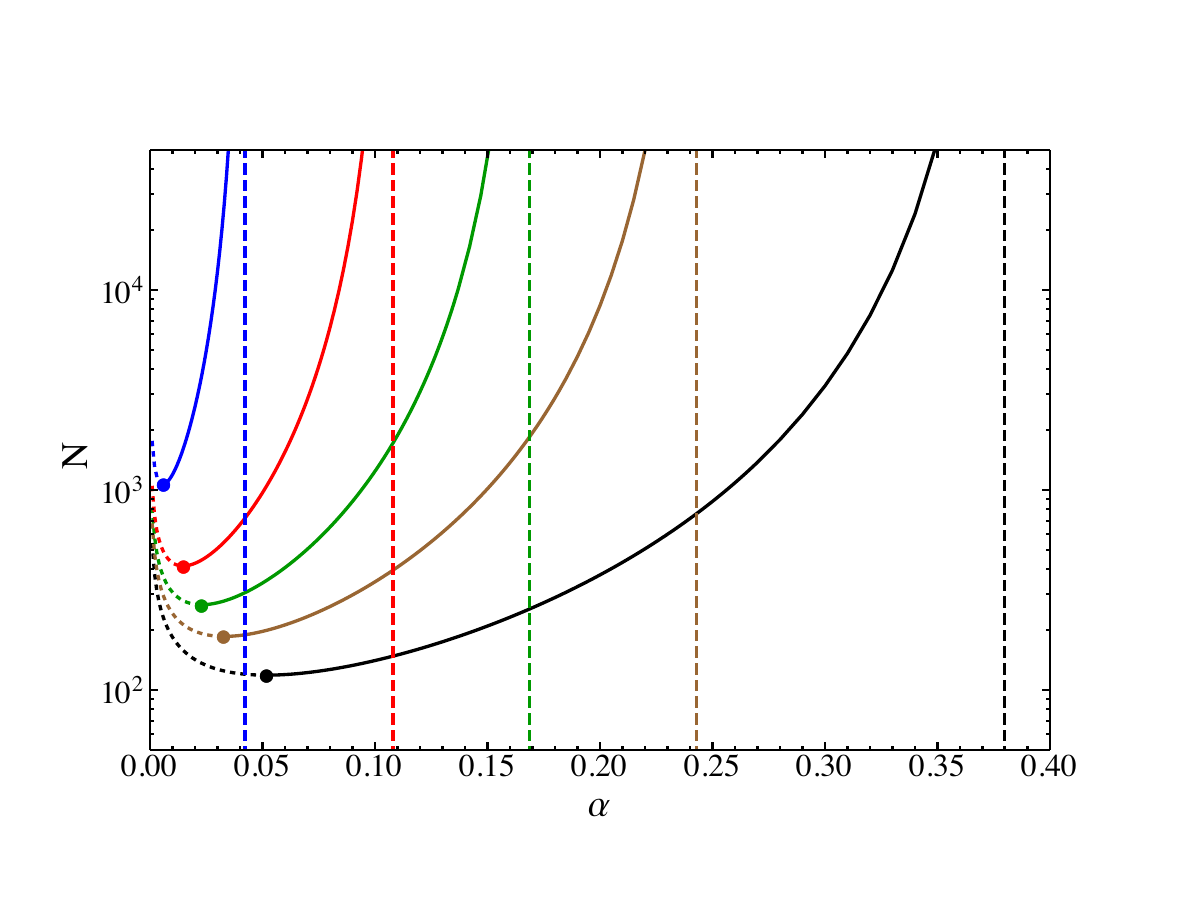}
    \caption{
   Schematic representation of the Vakhitov-Kolokolov stability criterion for two-field oscillons. We plot the quantity $N$, defined in Eq.~\eqref{eq:Ndef}, as a function of the phase-shift parameter $\alpha$ for $\Lambda=-0.5, -0.2, 0,0.2, 0.5$ (blue, red, green, brown and black	respectively). The vertical dashed lines correspond to the value of $\alpha_c$ for each case, while the solid (dotted) branches correspond to stable (unstable) oscillons.
 }
    \label{fig:VKrelations}
\end{figure}

\subsection{Oscillon Dynamics and Decay}

Having constructed two-field oscillon solutions and extended the V-K criterion to examine their stability, we move  to numerically demonstrate our results and  understand the relevant time-scales of interest. As a benchmark case of unstable oscillons, we use profiles corresponding to $\alpha=0.04 \alpha_c$ and $\epsilon=0.08$, for both attractive ($\Lambda = 0.5$) and repulsive ($\Lambda = -0.2$) interactions. which we perturb using the most unstable mode-functions computed in Section~\ref{sec:stability}. However, while the perturbation shape is given, the perturbation amplitude is a free parameter. We define the perturbation amplitude as the relative size of the oscillon and perturbation mode-function at the origin 
\beq
\delta_{\rm magn.} = {\delta(r=0,t=0) \over \phi_{\rm osc}(r=0,t=0)} \, ,
\eeq
and consider various values for $\delta_{\rm magn.}$. We investigate both the effect of placing the perturbation on both fields (with equal magnitude), and the effect of placing the perturbation on only one of the two fields and initializing the other field in the ``perfect'' oscillon configuration.

\begin{figure}[h]
    \centering
    \includegraphics[width=\textwidth]{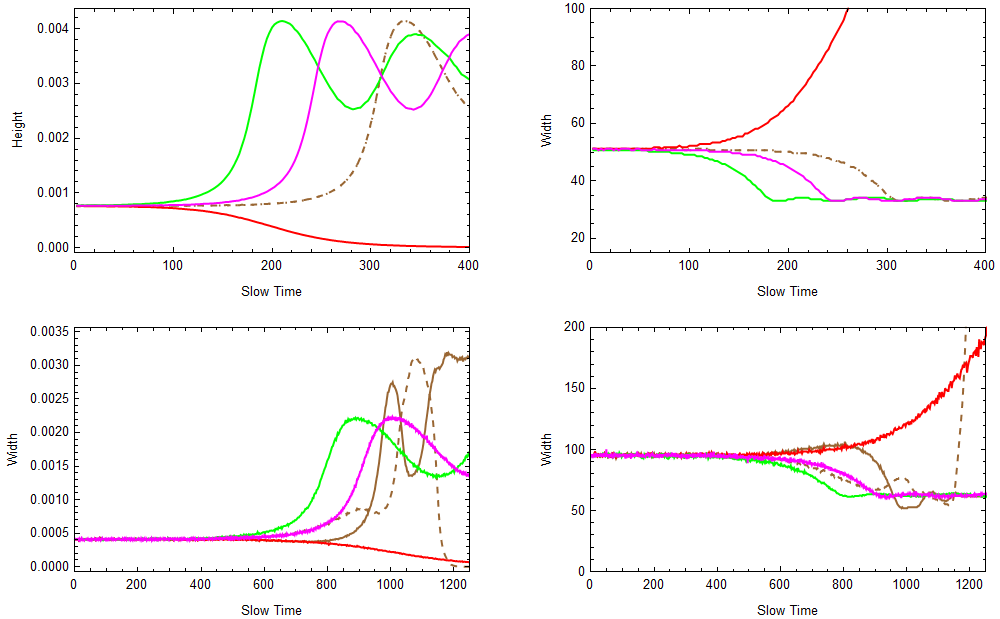}
    \caption{
\textit{Upper Panels:} The evolution of the oscillon peak energy density (left) and width (right) as a function of time $t (m^{-1})$ for $\Lambda=0.5$, $\alpha=0.04 \alpha_c$ and $\epsilon=0.08$. We perturb the oscillon by the most unstable mode with amplitude  $\delta_{\rm magn.}=-0.5\%, 0.05\%, 0.5\%$ (green, magenta and red respectively). The brown curve corresponds to perturbing only one of the two fields with $\delta_{\rm magn.}=0.2\%$ and initializing the other field as the ``perfect'' oscillon profile. The data for both fields is plotted for this curve (dashed and dotted line) but the difference is not perceivable since the fields quickly attract and undergo simultaneous collapse.
\textit{Lower Panels:} The same simulation for  $\Lambda=-0.2$ with again $\alpha = 0.04 \alpha_c$. Here the curves correspond to  $\delta_{\rm magn.}=-0.1\%, 0.01\%, 0.1\%$ (green, magenta and red respectively). We again plotted brown curves in a scenario where we only perturbed one of the fields with $\delta_{\rm magn.}=0.1\%$. The fate of the two oscillons differs as one of the field eventually disperses (dotted) while the other settles into a stable single-field oscillon (full). It is clear however that all of these multi-field oscillons contain unstable modes.}
   \label{fig:oscillondecay}
\end{figure}

Fig.~\ref{fig:oscillondecay} shows the time-evolution of the oscillon’s energy density at the origin and width as a function of slow time $\tau=\epsilon^2 t$. First of all, it is evident that the slow time controls the relevant time-scales for the oscillon dynamics. The time of decay of the oscillon seems to occur at a similar order of magnitude in slow time. This can be traced back to the evolution Eqs.~\eqref{eq:NLS} for the complex functions $A(\rho,\tau)$ and $B(\rho,\tau)$, which are of the non-linear Schr\"odinger type, and control the dynamics of the oscillon envelope, on top of the fast oscillations, which occur on a timescale of ${\cal O}(m^{-1})$.  
We see that the magnitude of the initial perturbation plays a crucial role for the ultimate fate of the oscillon: small fluctuations  $\delta_{\rm magn.}\ll 1\%$ lead to a new semi-stable point for the width and height of the localized configuration. On the other hand, for larger $\delta_{\rm magn.}$ the initial localized wave-packet simply disperses. 
We must note here, that initializing the two fields out of phase $a=-b$ and using the corresponding perturbation derived using Eqs.~\eqref{eq:linearper2} gives indistinguishable curves, and thus we do not plot them for clarity. This however confirms our analysis, that the  phase of the two fields in the initial  oscillon configuration, $0$ or $\pi$, does not affect the subsequent dynamics.
We will show numerically in Section~\ref{sec:emergence} that the phase relation between the components of the oscillon arises dynamically when they form, leading to phase-locked fields, oscillating either in or out of phase.

Furthermore, one can ask whether having the exact form of the most unstable modes computed in the previous section is a necessary condition for oscillon decay. In single-field oscillons, one can see that any perturbation can be decomposed in a basis of eigen-functions of the operator $H_1H_2$ [see Eqs.~\eqref{eq:A9} and \eqref{eq:A10}], meaning a generic perturbation will include a component along the unstable mode. In two-field systems, it is interesting to see how an initial perturbation is transferred from one field to the other.
To check the validity of our analysis, we perturbed only the $\phi$ field, leaving the $\chi$ field profile identical to the perfect oscillon solution. 
Perturbing one of the fields with $\delta_{\rm magn.}$ while leaving the other field intact leads in some cases to a similar evolution as perturbing both fields with $\delta_{\rm magn.}/2$.
For the specific cases that we plotted, this occurs for $\Lambda=0.5$ when perturbing one of the fields with $\delta_{\rm magn.}=0.2\%$, leading to almost indistinguishable evolution from perturbing both fields with  $\delta_{\rm magn.}=0.1\%$. We do not plot both these simulations in Fig.~\ref{fig:oscillondecay}, because the results are visually identical (instead only one of these cases is plotted).
 This indicates that due to the interaction of the two fields and the symmetry of the system, a small initial perturbation is quickly distributed evenly among the two fields.
 In other cases we  observed the complete dispersion of one field and the relaxation of the other towards a stable single-field oscillon solution (see lower panels of Fig.~\ref{fig:oscillondecay}).
 In particular this occurred for $\Lambda=-0.2$, when we perturbed only one of the two fields by $\delta_{\rm magn.}=0.1\%$. 
 From our numerical investigation it remains clear that the symmetric multi-field oscillons failing the VK-criterion contain unstable modes and will in general decay. However, when perturbed outside the $\phi=\pm \chi$ Ansatz, the dynamics can either restore or destroy the 
equality between the two fields. We will see that for $\Lambda>0$ the two-field configuration is a strong attractor of the system. 
We note that it is not necessary to perturb the oscillon with the exact form of the perturbation given by the stability analysis of this Section, since unstable oscillons will eventually decay if perturbed, albeit on slightly different time-scales and with different final state, depending on the details of the perturbation. Finally, we also perturbed the oscillon with negative $\delta_{\rm magn.}$ (meaning that the unstable mode is ``flipped''). We observed that this leads to a collapse instability more often than the positive $\delta_{\rm magn.}$ case. This highlights the sensitivity of the fate of the unstable oscillons on the initial conditions. Finally, we perturbed oscillons that are characterized as stable by the two-field version of the V-K criterion in various ways, seeing no decay over the total run of the simulation, lasting several hundred units of ``slow time'' $\tau = \epsilon^2 t$.

\begin{figure}
    \centering
    \includegraphics[width=0.8\textwidth]{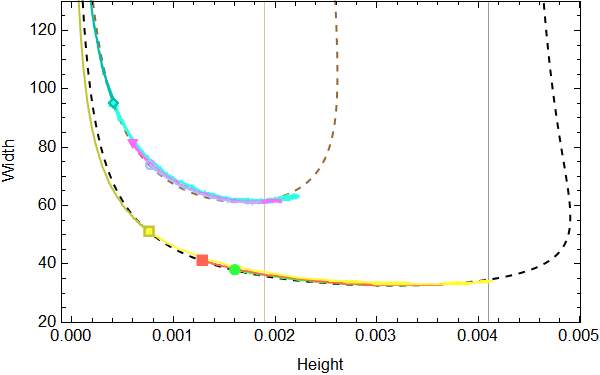}
    \caption{
The height and width of oscillons undergoing decay for $\Lambda=0.5$ and $\Lambda = -0.2$. These simulations were performed with $\epsilon = 0.08$. We plot the curves of oscillons undergoing collapse ($\delta_{magn} \ll 1\%$) where different colors correspond to different values of $\alpha$ with $\alpha = 0.04$ (cyan), $\alpha = 0.06$ (magenta), $\alpha = 0.08$ (purple) for $\Lambda = -0.2$ and $\alpha = 0.04$ (yellow), $\alpha = 0.07$ (red), $\alpha = 0.09$ (green) for $\Lambda = 0.5$. For $\alpha = 0.04$ we also plot curves corresponding to dispersion (dark yellow and dark cyan). Here $\alpha$ is given in terms of $\alpha_c$. We also plot the ``perfect" oscillon curves for $\Lambda = -0.2$ (brown) and $\Lambda = 0.5$ (black) and indicate the point of stability under VK with a gridline. It is clear that the oscillon configurations decay until they either disperse completely or find a new point of stability.
}   \label{fig:phasespace}
\end{figure}

It is now important to understand this behaviour. Fig.~\ref{fig:phasespace} shows the trajectories of several perturbed oscillons on the height-width curve for oscillons\footnote{In this plot, the height and width curve is derived from the energy density, not the profile of the oscillons. It is  thus  related to the one shown in Fig.~\ref{fig:heightwidth}, but is not identical to it.} for $\Lambda=0.5$ and $\Lambda = -0.2$. We see that unstable oscillons will surely move away from their initial configuration if perturbed. However their ultimate fate is not unique. It is known that localized solutions of the three-dimensional non-linear Schr\"odinger equation are unstable and  undergo a collapse instability, meaning that their width becomes smaller and their amplitude becomes larger.  For oscillons that are ``mildly perturbed'', $\delta_{\rm magn.}\ll 1\%$, this is exactly what we see at first: the unstable oscillons follow the height-width relation, on which their amplitude increases and their width decreases. This is the regime, where one can safely neglect the sextic term in the initial Lagrangian, equivalently the $|A|^4A$ term in Eq.~\eqref{eq:NLS}. Once the oscillon amplitude grows enough, so that the $|A|^4A$ term becomes important, the oscillon is stabilized, which occurs near the minimum of the height-width curve. Beyond this point, the oscillon solution  satisfies the V-K criterion. Because our simulation box does not allow for the energy to escape (it does not possess absorbing boundary conditions), the oscillon will never truly reach the ``ideal'' shape, and will instead stay in the vicinity of the first stable point on the height-width curve. 

Oscillons that are perturbed by a large amount, which puts them too far away from the height-width curve,  do not undergo this slow collapse instability, which would take them to a stable configuration. 
Using different perturbation types on the initial oscillon configuration amounts to examining
the basis of attraction for two-field oscillons, finding the approximate size of the initial perturbation below which the oscillon is allowed to ``rearrange'' itself into a stable configuration and above which the oscillon is completely destabilized.
An important question remains, as to the oscillon configurations that arise naturally after inflation: it is not known if a universe that is governed by the Lagrangian of Eq.~\eqref{eq:Action} during preheating will be dominated by single or multi-field oscillons. We are currently working on   the corresponding three dimensional lattice simulations required to address this point and will present the results in a subsequent publication.

\section{Oscillon lifetime}
\label{sec:lifetime}

It is a well known fact that oscillons are not exactly stable, but eventually lose their energy through an exponentially suppressed radiative tail \cite{Fodor_2009, Fodor_2009-2}. To make an assessment of the (cosmological) importance of the stable oscillons that were found in the previous sections we wish to find an estimate of their lifetimes. To do this we follow the analyses performed in Refs.~\cite{Zhang_2020classdec, Ibe_2019, Mukaida_2017}.
 The core reason why oscillon solutions are in general not stable (classically), even though they are long-lived, is that plugging in an oscillating solution $\sim \Phi_0 \cos(\omega t)$ into an equation of motion with nonlinear terms will in general lead to terms that are proportional to higher harmonics $\propto \cos(j \omega t)$ with $j > 1$. In general, these terms can be made to cancel out using some perturbative procedure (e.g.~the two-timing analysis); but since in practice these perturbative techniques must end at some finite order and do not converge, there will always be residual terms of higher harmonics in the equations of motion. These terms will act as a source for radiative modes of the oscillon solution.\footnote{A class of arbitrarily long-lived oscillons was recently discussed in Ref.~\cite{Olle:2020qqy}.} In order to estimate the decay rate of the oscillons constructed in Section~\ref{sec:expansion} we  follow the methods of Refs.~\cite{Zhang_2020classdec, Ibe_2019, Mukaida_2017}.
To study the relevant radiative modes we perturb our oscillon solutions $\phi= \phi_{osc} + \delta$, $\chi = \chi_{osc} + \xi$, plug these into the equations of motion and linearize to obtain
\beq
\begin{split}
\left[ \partial_t^2 \phi_{osc} - \nabla^2 \phi_{osc} + V_{,\phi}(\phi_{osc}, \chi_{osc})\right] + \left[ \partial_t^2 \delta - \nabla^2 \delta + V_{,\phi \phi}(\phi_{osc}, \chi_{osc}) \delta +  V_{,\phi \chi}(\phi_{osc}, \chi_{osc}) \xi \right] = 0
\\
\left[ \partial_t^2 \chi_{osc} - \nabla^2 \chi_{osc} + V_{,\chi}(\chi_{osc}, \phi_{osc})\right] + \left[ \partial_t^2 \xi - \nabla^2 \xi + V_{,\chi\chi}(\chi_{osc}, \phi_{osc}) \xi +  V_{,\chi \phi}(\chi_{osc},\phi_{osc}) \delta \right] = 0
\end{split}
\label{eq:perturbedeqmot}
\eeq

\begin{figure}[h]
    \centering
    \includegraphics[width=1\textwidth]{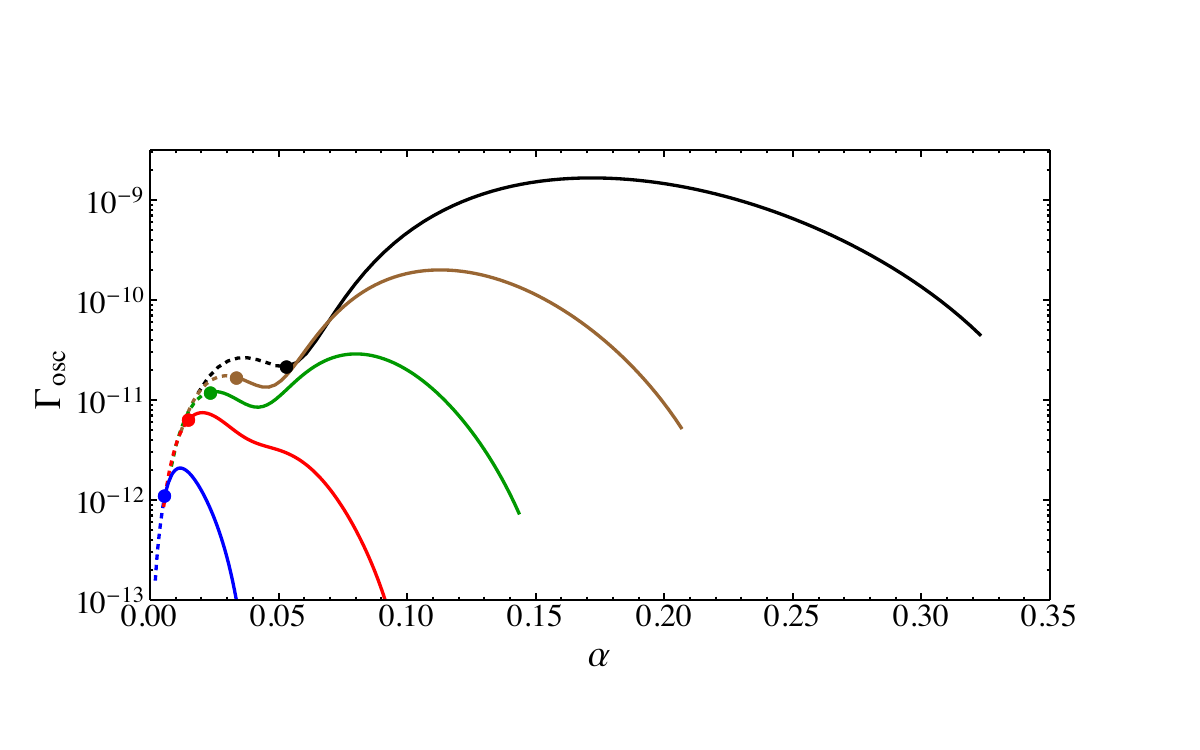}
    \caption{
The decay rate of oscillons computed computed using Eq.~\eqref{eq:decayrate} for $\epsilon=0.1$, as a function of the frequency parameter $\alpha$ for $\Lambda=-0.5, -0.2, 0,0.2, 0.5$ (blue, red, green, brown and black	respectively). The solid (dotted) branches correspond to stable (unstable) oscillons according to the VK criterion.
}  
 \label{fig:decayrate}
\end{figure}

In the above equations we use the notation $V_{,\phi}(\phi_{osc}, \chi_{osc}) \equiv \left .\partial_\phi V(\phi,\chi)\right |_{\phi \to \phi_{\rm osc}, \chi\to \chi_{osc}}$ for derivatives of the potential with respect to the fields.
We now  plug in our oscillon solutions $\phi_{\rm osc} = \Phi_0 \cos(\omega t)$ and $\chi_{\rm osc} \sim X_0 \cos(\omega t)$, where $\omega = 1 -\alpha \epsilon^2/2$. By virtue of the two-timing analysis, all terms proportional to $\cos(\omega t)$ in the first square bracket of Eqs.~\eqref{eq:perturbedeqmot} cancel and we obtain
\beq
\begin{split}
\partial_t^2 \delta - \nabla^2 \delta + \delta - \left(3 \phi_{osc}^2 - 5 g \phi_{osc}^4 + \Lambda \chi_{osc}^2\right) \delta - 2\Lambda \phi_{osc} \chi_{osc} \xi =&
\\
 \left(\frac{1}{4}\Phi_0^3 + \frac{1}{4}\Lambda \Phi_0 X_0^2 - \frac{5}{16} g \Phi_0^5\right)\cos(3 \omega t) &- \frac{1}{16} g \Phi_0^5 \cos(5\omega t)
\\
\partial_t^2 \xi - \nabla^2 \xi + \xi - \left(3 \chi_{osc}^2 - 5 g \chi_{osc}^4 + \Lambda \phi_{osc}^2\right) \xi - 2\Lambda \phi_{osc} \chi_{osc} \delta = &
\\ \left(\frac{1}{4}X_0^3 + \frac{1}{4}\Lambda X_0 \Phi_0^2 - \frac{5}{16} g X_0^5\right)\cos(3 \omega t) & - \frac{1}{16} g X_0^5 \cos(5\omega t)
\end{split}
\label{eq:rad}
\eeq
Focusing on the solutions where the two fields within the oscillon oscillate in phase $\Phi_0 = X_0$ (the  case where $\Phi_0 = - X_0$ is analogous); we add Eqs.~\eqref{eq:rad} and introduce the new radiative mode $\Psi = \delta + \xi$ (in the case where the oscillons oscillate out of phase we would introduce $\Psi = \delta - \xi$
) to obtain the single equation
\beq
\partial_t^2 \Psi - \nabla^2 \Psi + \Psi - \left(3(1 + \Lambda) \phi_{osc}^2 - 5 g \phi_{osc}^4\right) \Psi = \left(\frac{1}{2}\Phi_0^3 + \frac{1}{2}\Lambda  \Phi_0^3 - \frac{5}{8} g \Phi_0^5\right)\cos(3 \omega t) - \frac{1}{8} g \Phi_0^5 \cos(5\omega t)
\label{eq:radsingle}
\eeq
We are thus interested in wave-like solutions $\Psi_{rad}(r, t)$ of Eq.~\eqref{eq:radsingle}. Using the symmetry of our system we can then deduce that half the amplitude of this wave comes from the oscillon in the $\phi$-field and the other half from the oscillon in the $\chi$-field. Notice however that this calculation is equivalent to calculating the radiative mode of an oscillon in a single field with a slightly altered quartic term in the potential
\beq
V(\phi) = \frac{1}{2} \phi^2 - \frac{1}{4}(1 + \Lambda) \phi^4 + \frac{g}{6} \phi^6  \, .
\label{eq:potaltered}
\eeq
As suggested in Ref.~\cite{Zhang_2020classdec}, when calculating the outgoing radiation of an oscillon  in this type of potential, ignoring the spacetime-dependent effective mass term in the equation of motion for the perturbative mode will not alter the final result significantly. Thus we  ignore the time-dependent coupling of $\Psi$ on the left-hand-side of Eq.~\eqref{eq:radsingle} and obtain the simpler equation
\beq
\partial_t^2 \Psi - \nabla^2 \Psi + \Psi = \left(\frac{1}{2}\Phi_0^3 + \frac{1}{2}\Lambda  \Phi_0^3 - \frac{5}{8} g \Phi_0^5\right)\cos(3 \omega t) - \frac{1}{8} g \Phi_0^5 \cos(5\omega t)
\, .
\label{eq:radsinglesimple}
\eeq
As was shown in Ref.~\cite{Zhang_2020classdec}, this equation can be solved by expanding $\Psi(r, t) = \sum_{j = 1}^{\infty}\Psi_j(r)\cos(j\omega t)$. Obviously only $\Psi_3(r)$ and $\Psi_5(r)$ will not vanish in this specific case. 
The rate of energy loss of the oscillon is $dE/dt = 4\pi r^2 T_{0r}$, where $T_{0r}$ is the Poynting vector in our spherically symmetric set-up. By averaging over time in the far distance regime 
\beq
\langle T_{0r} \rangle = \langle  \dot \xi \partial_r \xi \rangle  +\langle  \dot \delta \partial_r \delta \rangle = \frac{1}{2} \langle  \dot \Psi \partial_r \Psi \rangle \, .
\eeq
Following Ref.~\cite{Zhang_2020classdec} we obtain the following analytic expression for the outgoing radiation at large radii
\beq
\Psi_{rad}(r, t) \approx \frac{1}{4 \pi r}\left(S_3(k_3)\cos(k_3 r - 3 \omega t) + S_5(k_5)\cos(k_5 r - 5 \omega t)\right) \, ,
\label{eq:radwave}
\eeq
where $k_j = \sqrt{(j\omega)^2 - 1}$ and $S_3(k)$ and $S_5(k)$ are the Fourier transforms of the terms on the right hand side of Eq.~\eqref{eq:radsinglesimple} proportional to the third and fifth harmonic respectively. Thus
\beq
S_3(k) = \int d^3r \left(\frac{1}{2}\Phi_0^3 + \frac{1}{2}\Lambda  \Phi_0^3 - \frac{5}{8} g \Phi_0^5\right) e^{-ik \cdot r} = \int_0^\infty dr 4\pi r \frac{\sin(kr)}{k} \left(\frac{1+ \Lambda}{2}\Phi_0^3  - \frac{5}{8} g \Phi_0^5\right)
\eeq
and
\beq
S_5(k) =- \int d^3r \frac{1}{8} g \Phi_0^5  e^{-ik \cdot r} = -\int_0^\infty dr 4\pi r \frac{\sin(kr)}{k} \frac{1}{8} g \Phi_0^5  \, .
\eeq
We find, similarly to Ref.~\cite{Ibe:2019vyo}, that $S_3\gg S_5$.
The analytic expression for the radiative modes of the oscillon given by Eq.~\eqref{eq:radwave} leads to the following expression for the decay rate of the oscillon
\beq
\Gamma_{osc} = \frac{1}{16 \pi E_{osc}}\sum_{j=3,5} S_j(k_j)^2 \omega_j k_j
\, .
\label{eq:decayrate}
\eeq
By using the oscillon amplitude   $\Phi_0 \equiv \epsilon a(\epsilon r)$, which corresponds to the solution of  Eq.~\eqref{eq:profs}, we computed the value of $\Gamma_{osc}$  using the method described above. 
We confirmed that the multi-peak structure of $\Gamma$ found in Ref.~\cite{Ibe:2019vyo} persists for different values of $\Lambda$, either positive or negative. 
In general, the oscillon lifetime increases as one decreases the small parameter $\epsilon$. In our construction, this corresponds to increasing the strength of the sextic term. 
Fig.~\ref{fig:decayrate} shows the results derived by numerically integrating Eq.~\eqref{eq:decayrate} for three-dimensional oscillons, using the value $\epsilon=0.1$. For  $\Lambda=0$ we recover the results of Ref.~\cite{Kasuya:2002zs}.
We see that the decay rate increases for $\Lambda>0$ compared to the single field case ($\Lambda=0$) and correspondingly it decreases for $\Lambda<0$. The maximum value of $\Gamma$ increases from $\Lambda=0$ to $\Lambda=0.2$ and $\Lambda=0.2$ to $\Lambda=0.5$   by around one order of magnitude.

\section{Generalization to many fields}
\label{sec:Nfields}
In this Section we  generalize our model to a system of $N$ interacting fields. 
Following the two-field  analysis, we choose the following action where the fields exhibit an ``exchange'' symmetry:
\beq
S =- \int d^3x dt \left [
\sum^N_{I=1} \left ( {1\over 2} \partial_\mu \phi_I\partial^\mu \phi_I + {1\over 2} m^2 \left (\phi_I \right )^2
- {\lambda\over 4} \left  (\phi_I \right )^4 + \frac{g}{6}\left (\phi_I \right  )^6
\right ) -\sum_{I \neq J} \frac{\Lambda}{2}(\phi_I)^2 (\phi_J)^2
\right ]
\, .
\label{eq:Actionnfields}
\eeq
We  briefly repeat the derivations presented above for two-field oscillons, showing that that the profile equations are solved by zero-modes of Eq.~\eqref{eq:profs} with an altered cubic term. Furthermore, we show how to extend the V-K criterion to apply to the $N$-field system.

\subsection{Two-timing analysis}
After rescaling the fields and spatial-temporal dimensions in the same way as was done in section \ref{sec:Model}, the action of Eq.~\eqref{eq:Actionnfields} results in $N$ equations of motion for the $N$ fields. The equation for the $i$-th field (assuming spherical symmetry) is
\beq
\partial_t^2 \phi_i - (\partial_r^2 + \frac{2}{r}\partial_r) \phi_i + \phi_i = \phi_i^3 - g \phi_i ^5 + \sum_{j\neq i} \Lambda \phi_i \phi^2_j
\label{eq:motionN}
\eeq
Introducing the by now familiar two-timing analysis variables $\rho = \epsilon r$ and $\tau =\alpha \epsilon^2 t$, the equations become
\beq
\partial_t^2\phi_i + 2\alpha\epsilon^2 \partial_t \partial_{\tau} \phi_i  - \epsilon^2(\partial_{\rho}^2 + \frac{2}{\rho}\partial_{\rho}) \phi_i + \phi_i = \phi_i^3 - g \phi_i ^5 + \sum_{j\neq i} \Lambda \phi_i \phi^2_j+ {\cal O}(\epsilon^4)
\label{eq:motionNeps}
\eeq
Finally looking for solutions $\phi_i = \epsilon \phi_{1,i} + \epsilon^2 \phi_{2,i} + ...$ we can write down the $O(\epsilon)$ and $O(\epsilon^3)$ equations
\begin{eqnarray}
{\cal O}(\epsilon)&:& ~ \partial_t^2 \phi_{0,i} + \phi_{0,i} = 0
\\
{\cal O}(\epsilon^3)&:& ~
\partial_t^2 \phi_{2,i} + \phi_{2,i} = (\partial_{\rho}^2 + \frac{2}{\rho}\partial_{\rho})\phi_{0, i} - 2\alpha \partial_t \partial_{\tau} \phi_{0, i} + \phi^3_{0, i} -  \phi^5_{0, i}  + \sum_{j\neq i} \Lambda \phi_{0, i} \phi^2_{0, j}
 \, ,
\end{eqnarray}
where we've again used $g = 1/\epsilon^2$. The rest of the derivation is entirely analogous to the one in Section~\ref{sec:expansion} so we won't repeat it here. The procedure results in $N$ profile equations 
\beq
(\partial_{\rho}^2+\frac{2}{\rho}\partial_{\rho}) a_i - \alpha a_i + \frac{3}{4}a_i^3 - \frac{5}{8}a_i^5 + \frac{3}{4}\Lambda \sum_{j \neq i} a_i a_j^2 = 0
\label{eq:profilesN}
\eeq
We thus need to find a set of $N$ functions $a_i(\rho)$ that solve this system of equations. As before, the symmetry in the model greatly simplifies this question. We need to find localized solutions of the equation
\beq
(\partial_{\rho}^2+\frac{2}{\rho}\partial_{\rho}) a - \alpha a + \frac{3}{4}(1 + \Lambda(N - 1))a^3 - \frac{5}{8}a^5 = 0
\label{eq:profilessimpleN}
\eeq
Then, setting any $n < N$ number of $a_i(\rho) = a(\rho)$ and the remaining $a_j(\rho) = -a(\rho)$ solves the system in Eq.~\eqref{eq:profilesN}. 
Notice that Eq.~\eqref{eq:profilessimpleN} has the same form as Eq.~\eqref{eq:profs} with an altered cubic term. It will therefore have localized solutions, which are related to the ones for the two-field model by substituting the coupling strength $\Lambda$ in the two-field case with the ``effective'' coupling strength  $\Lambda(N-1)$ in the multi-field case. This is a proof of existence for such a solution, not uniqueness.

\subsection{Stability analysis}
For brevity, we move the full derivation of the stability of these $N$-field oscillons to long wavelength perturbations to  Appendix~\ref{app:VK}. Note however that performing the two-timing analysis and using the ansatz $\phi_{0, i} = \operatorname{Re}\left \{A_i(x, t)e^{-it}\right\}$ is equivalent to working in the nonrelativistic limit. In this limit, the Lagrangian has a conserved charge proportional to
\beq
\mathcal{Q} = i \int d^dx \sum_i\left(A^*_i\dot{(A_i)} - A_i\dot{(A^*_i)}\right) \, ,
\eeq
which reduces to the following after inserting the symmetric oscillon solution
\beq
\mathcal{Q} =N_{fields}\cdot \alpha \epsilon^2 \int d^dx \cdot a(x)^2 \, .
\eeq
Since the oscillon has a conserved charge, we can apply the VK-criterion directly. To summarize, the oscillon is stable when $\frac{\partial \mathcal{Q}}{\partial \alpha \epsilon^2} > 0$. This is just an alternate and equivalent form of the VK-criterion that was given earlier in Eq.~\eqref{eq:Ndef}. The criterion for stability therefore still holds after a generalization of our model to $N$ fields.\\\\
To conclude, we constructed  oscillons in a model with $N$ fields. Since we work in the nonrelativistic limit it is simple to show that the VK criterion still applies. Although we have not performed a detailed analysis of the lifetime of these oscillons, the result is expected to be analogous to the two-field case. Namely, the computation of the radiation tail of these oscillons can be mapped to an equivalent calculation for single-field oscillons with an altered cubic term in the potential, as done in Section~\ref{sec:lifetime} for the two-field case.

\section{Oscillon Emergence and dynamics}
\label{sec:emergence}

Despite the mathematical construction and stability analysis of two-field oscillons, the question of their emergence in realistic scenarios and hence their cosmological consequences remains. In this section we provide a combination of simulations in one and three spatial dimensions, in both an expanding and a static background. The full three-dimensional simulations of oscillon formation after inflation and their GW signatures will be presented in a subsequent publication.

\subsection{Dynamical emergence in 1+1D}
\label{sec:expanding}

To investigate the formation of oscillons in this model we performed a series of one-dimensional simulations of preheating-like scenarios on an expanding background. 
One of the fields, $\phi$, was initialized as a homogeneous condensate, akin to the inflaton at the end of inflation. The other fields of the model $\chi, \theta,...$ were then initiated in their vacuum, akin to spectator fields at the end of inflation. The simulations were strongly inspired by those performed in Ref.~\cite{amin:1demergence}.
We see a clear sequence of events, leading from an oscillating $\phi$ condensate to a collection of composite oscillons.
\begin{itemize}
\item Parametric self-resonance leads to a fragmentation of the inflaton condensate and the formation of single-field oscillons.
\item Parametric resonance of the spectator field(s) by the single-field oscillons leads to exponential growth of the spectator, akin to preheating and  Floquet theory.
\item Once the spectator field acquires a large enough value, non-linear interactions between the two (or more) fields lead to phase-locked long-lived configurations: composite oscillons.
\end{itemize}

\begin{figure}[h!]
    \centering
    \includegraphics[width=.45\textwidth]{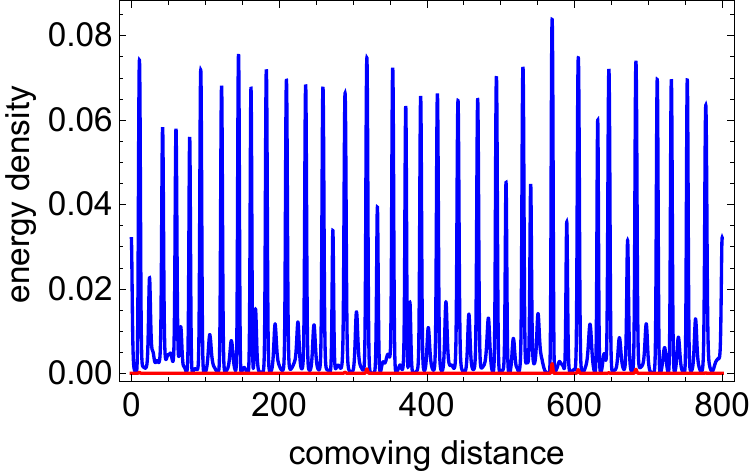}    \includegraphics[width=.45\textwidth]{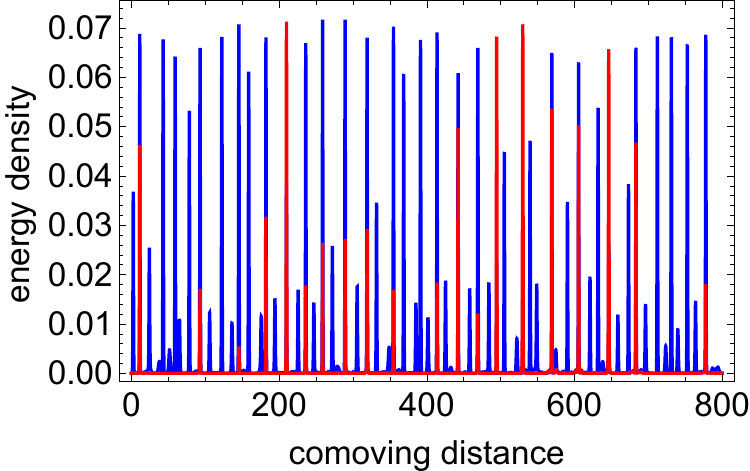}
    \caption{
    The energy density in the box for a two-field simulation, involving the fields  $\phi$ and $\chi$  (blue and red respectively) for $t = 1750$ $(m^{-1})$ (left) and $t=4500$ $(m^{-1})$  (right). The $\phi$ field starts with fluctuations on top of a  classical homogeneous configuration (emulating the inflaton), while the the $\chi$ field is comprised solely of fluctuations initially.}
    \label{fig:box1D}

\end{figure}

\begin{figure}[h!]
    \centering
    \includegraphics[width=1\textwidth]{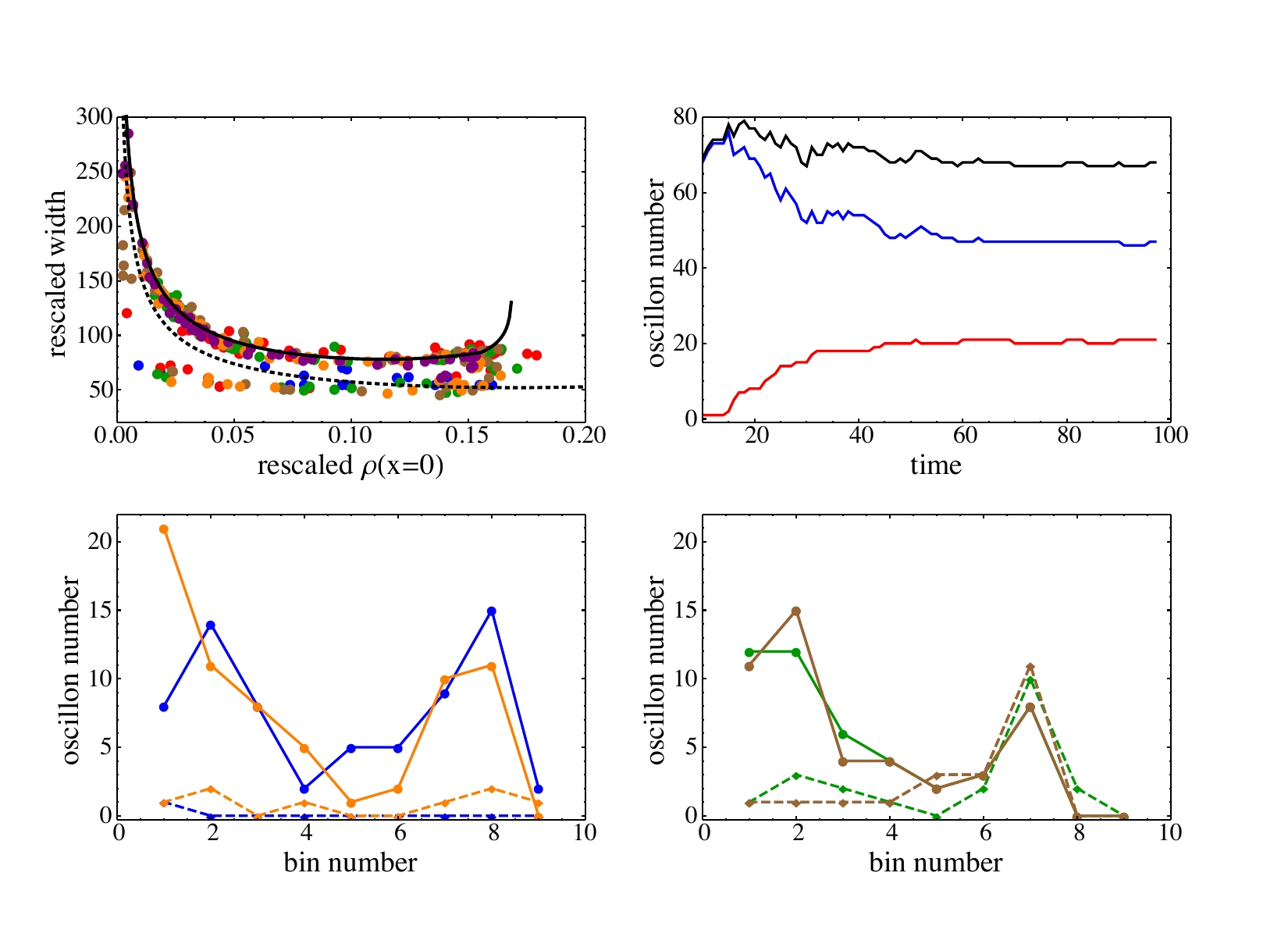}    

    \caption{
  {\it Upper left:} The height width relation for overdensities in the $\phi$  field  for $t=1500, 2500, 3300, 4000, 5200, 6700$ $(m^{-1})$ (purple, brown, blue, red, green). The solid (dashed) line corresponds to the $\Lambda=0$ ($\Lambda=0.5$) analytical profiles.
    {\it Upper right:} The number of single- and double-component oscillons (blue and red respectively). The total number is in black.
    {\it Lower panels:} The number of one- and two-component oscillons (solid and dashed lines respectively) in equal bins of height for $t=1500,25000,5200,6700$ $(m^{-1})$ (blue, orange, green and brown respectively).
 }
    \label{fig:twofieldlattice}
\end{figure}

Fig.~\ref{fig:box1D} shows the energy density on the one-dimensional lattice at early and late times. The oscillons appear to become narrower, because we plot our data on a comoving grid.  
Furthermore, we clearly see the above stages of oscillon formation and evolution. 
In the left panel, the $\phi$ condensate has fragmented into localized configurations (oscillons), while the $\chi$ field remains small; we initialized the $\chi$ field with $\langle \chi_k^2\rangle ={\cal O}(1/2k)$.
The   right panel clearly shows spikes of the $\chi$ field, positioned at the same locations as $\phi$ field oscillons. Interestingly, not all $\phi$ oscillons lead to an efficient amplification of the $\chi$ field. Furthermore, we see that the energy density spikes of the two fields are not equal in size. In fact, two-field composite oscillons in one spatial dimension exhibit a persistent energy exchange between the two fields, which we will describe more later. 


Fig.~\ref{fig:twofieldlattice} shows the characteristics of the oscillons in the one-dimensional simulation. In the top left panel we see the distribution of $\phi$ overdensities on the height-width plane. The formation of two populations is evident. A large number of overdensities closely follows the single-oscillon height-width curve. Furthermore, the early-time population of overdensities is largely or solely on this curve. For later times, we see a splitting between the single and two-field oscillon curves. The fact that the oscillons who deviate from the single-field curve do not follow exactly the two-field one is explained by the presence of a ``breathing'' mode between the two fields (see Section~\ref{sec:1Dfloquet}). The final number of two-field oscillons is almost half that of their single-field counterparts. 
The lower panels of Fig~\ref{fig:twofieldlattice} show the distribution of oscillons in equal bins of core energy density (amplitude). The left panel shows early times, with the orange curve corresponding to the initial stages of the emergence of $\phi-\chi$ oscillons. For later times (right panel) we see that the distribution is largely constant. Due to the finite bin size and the oscillon breathing mode, some oscillons will inevitably enter and exit adjacent bins, leading to small changes. However, the overall shape is indeed stable, allowing us to reach significant conclusions. Large amplitude oscillons are comprised of almost equal populations of one- and two-component configurations. Small amplitude oscillons on the other hand are almost entirely single-field. This can be easily understood as follows: Parametric resonance for $\chi$ is strongly sensitive to the amplitude of the pumping field $\phi$. It is thus much more efficient for the $\chi$ field to be excited in regions of large $\phi$ field, leading to a more efficient production of large field two-component oscillons. We will return to this point in Sections~\ref{sec:1Dfloquet}, \ref{sec:3Dfloquet} and \ref{sec:floquet}.
Finally, we checked the relative phase of the two fields in the two-field oscillons and  found no preference for either in-phase or out-of-phase configurations.

\begin{figure}[h!]
    \centering
    \includegraphics[width=1\textwidth]{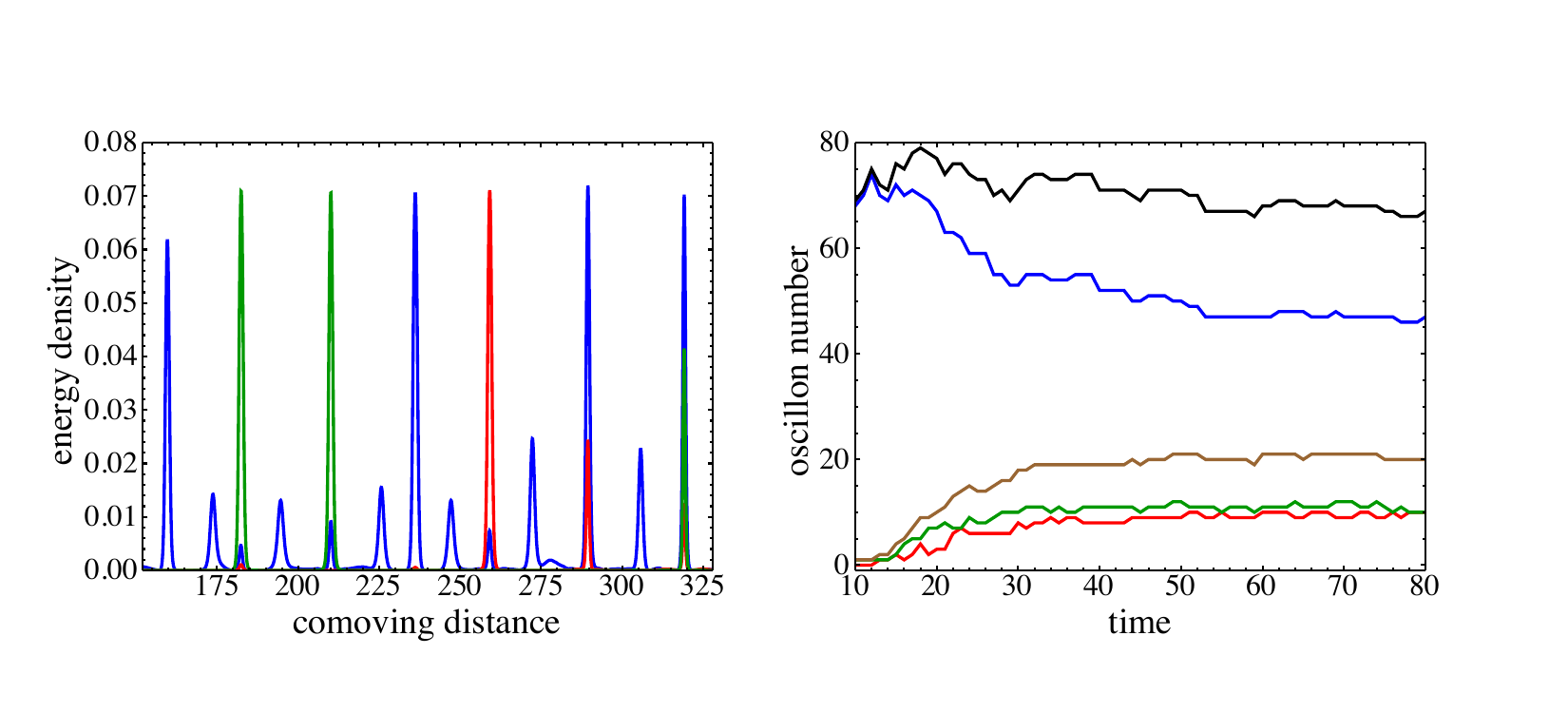}    

    \caption{
  {\it  Left:} The energy density in a zoomed-in part of the  box for a three-field simulation, involving the fields  $\phi$, $\chi$ and $\theta$  (blue and red and green respectively) for $t= 3300$ $(m^{-1})$. The $\phi$ field starts with fluctuations on top of a  classical solution (emulating the inflaton), while the the $\chi$ and $\theta$ fields are comprised solely of fluctuations initially. 
    {\it  Right:} The number of one-, two-component and three-component oscillons (blue, red and green respectively). The sum of two- and three-component oscillon is shown in brown and the total number of oscillons in black.
 }
    \label{fig:threefieldlattice}
\end{figure}

Fig.~\ref{fig:threefieldlattice} shows the formation of multi-component oscillons in a simulation involving three scalar fields. 
The system exhibits a similar evolution as in the two field case, with the $\phi$ fragmentation followed by single-field oscillon formation, which then leads to excitation of the spectator fields and the eventual emergence of composite oscillons. As in the two field case, not all single field oscillons lead to multi-field configurations. Interestingly the number of two- and three-component oscillons is nearly the same and the total number of multi-component oscillons is nearly equal to the number of two-component oscillons that emerged in the two-field simulation.

\subsection{Single to multi-field oscillons in  1+1D}
\label{sec:1Dfloquet}

The results of Section~\ref{sec:expanding} showcased the two-stage formation of composite oscillons on an expanding background for the one-dimensional case.
In order to better understand the transition from single to multi-field oscillons,  we focus on simulating single-field oscillons on a static background, which allows us to fully control the initial conditions and compare the numerical results to theoretical predictions. The decay into a multi-field configuration can be understood in the context of Floquet theory. Although obtaining analytical results can be challenging in a non-homogeneous background such as an oscillon, we show in Section~\ref{sec:floquet}   that significant intuition is gained by working in the homogeneous limit.

We initialize the $\phi$ field as an oscillon that is stable in the single field theory (with $\Lambda = 0$) and draw the fluctuations of the $\chi$ field from a Gaussian distribution with $\langle\chi_k^2\rangle = 1/2k$, as in our lattice simulations. Alternatively, initializing the secondary field as a small perturbation of the same shape as the oscillon does not change the outcome of the simulation (although it somewhat accelerates the decay into a multi-field configuration). We will demonstrate this last point later on in this section. 

\begin{figure}
    \centering
   \includegraphics[width = .45\textwidth]{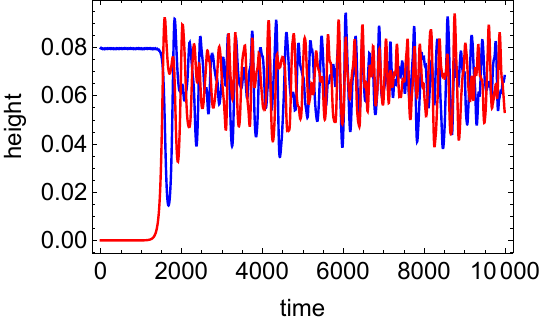}
   ~~~
    \includegraphics[width = .45\textwidth]{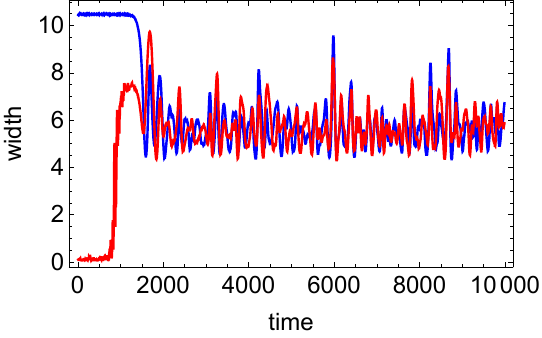}
   \\
    \includegraphics[width = .45\textwidth]{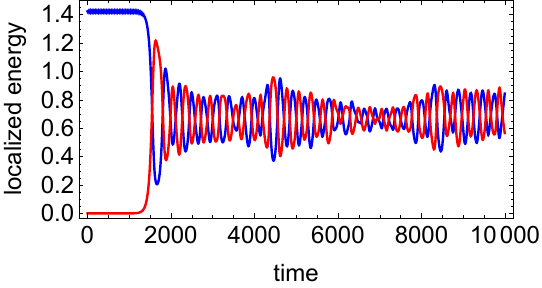}
   ~~~
    \includegraphics[width = .45\textwidth]{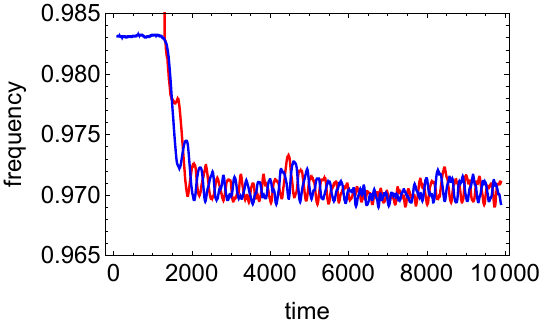}
   \\
    \includegraphics[width = .45\textwidth]{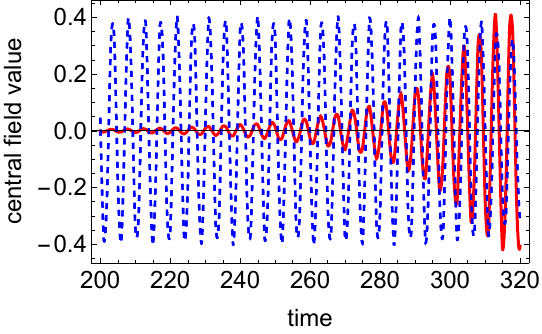}
   ~~~
    \includegraphics[width = .45\textwidth]{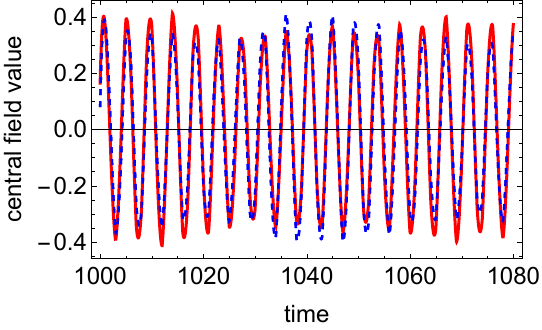}
    \caption{The evolution of various quantities in our simulations for the $\phi$ field (blue) and $\chi$ field (red). \textit{Top left:} the evolution of the height of the energy density at the origin. \textit{Top right:} the evolution of the width of the two fields defined as the distance where the energy density has decreased by a factor of $1/e$. \textit{Middle left:} the evolution of the localized energy in the two fields. \textit{Middle right:}
The oscillation frequency of the two fields.
\textit{Bottom panels:}
    The amplitude of the fields at the origin at early (left) and late times (right). It is clear that the single-field oscillon evolves into a two-field configuration, where the fields are phase locked (oscillating in phase) and share common characteristics.}
    \label{fig:1devolution}
\end{figure}

Fig.~\ref{fig:1devolution} shows the evolution of the system for $\Lambda = 0.5$ and $g = 1/\epsilon^2 = 5$: from a single field oscillon with $\alpha = 0.99 \alpha_c$ to a phase-locked two-field configuration.
The upper panels show the evolution of the width and height (in energy density space) of the two fields. We clearly see the growth of a localized $\chi$ configuration, which backreacts on the $\phi$-oscillon, leading to a composite oscillon. We see that after $t\simeq 1500 (m^{-1})$ the amplitude and width of the two fields oscillate around an average value, which is common for both. Inspecting the data we can postulate the existence of ``breather'' modes: stable perturbations atop the perfect oscillon solution that oscillate between the two fields (similarly to how we construct  unstable modes in App.~\ref{app:VK}). It seems that fewer of such breather modes  exist in higher dimensions (see Section~\ref{sec:3Dfloquet}).\\
The oscillation frequency behaves similarly. Initially, since the $\chi$ field is comprised of random fluctuations, one cannot speak of a single ``coherent'' frequency. As instabilities build up over time however, the two fields "lock" up into a single shared frequency as can be seen in the bottom  panels. The localized energy in each field also exhibits strong oscillations of up to $40\%$ around the average value. While the two components of the oscillon exchange energy with each other, the total configuration is still stable. This is strongly reminiscent of the behavior exhibited by oscillons in the $SU(2)$ gauged Higgs model \cite{Sfakianakis:2012bq}, where stable two-field oscillons discovered in the reduced spherical Ansatz exhibit similar  exchange of energy with a period much larger than the natural period of oscillation of the fields themselves.

Before concluding this section, we would like to explore the dependence of the evolution on the initial conditions for $\chi$, by altering them in two ways:  by drawing a different sample from the same aforementioned Gaussian distribution on the one hand and by initializing $\chi$ as a small ``bump'' with similar width to the $\phi$ oscillon, but very small amplitude.

\begin{figure}
    \centering
    \includegraphics[width = \textwidth]{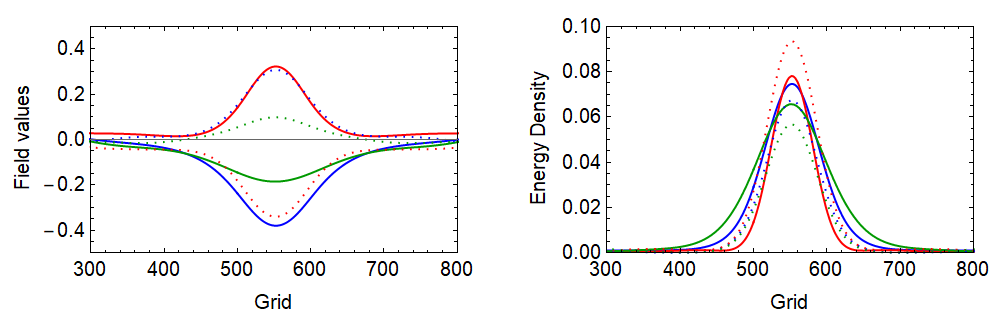}
    \caption{Time slices of the fields $\phi$ (full) and $\chi$ (dotted) at different times $t = 1900, 6000, 8000 (m^{-1})$ (blue, red and green respectively). The parameters are the same as those in Fig.~\ref{fig:1devolution}, but with a different Gaussian initialization for the field $\chi$. Contrary to Fig.~\ref{fig:1devolution} the fields condense into an out-of-phase configuration.}
    \label{fig:outofphase}
\end{figure}

We see in Fig.~\ref{fig:outofphase} that drawing a different sample from the same Gaussian distribution leads to a similar evolution with a phase-locked final state, albeit with the fields oscillating out of phase (in contrast to the situation of Fig.~\ref{fig:1devolution}). This verifies our assumption in the analytical construction of the oscillon, where we distinguished between in and out of phase oscillons, while at the same time showing that these two phase configurations are the attractors of the system, depending only on the initial conditions.

\begin{figure}
    \centering
    \includegraphics[width = \textwidth]{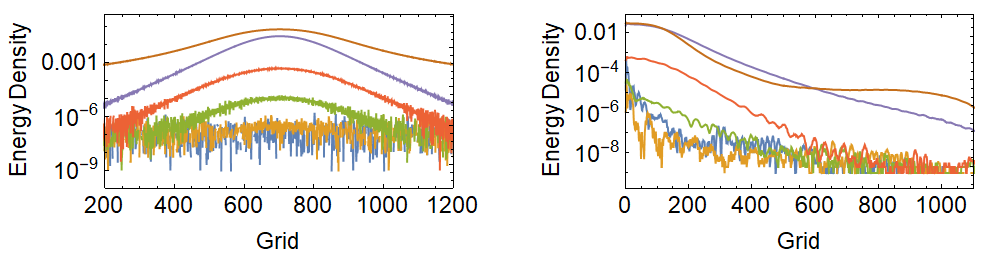}
    \caption{The evolution of the energy density  (time moves from bottom to top) of the secondary field starting from Gaussian initial conditions. It is clear that, although the initial configuration is very noisy, the field aligns into a smooth configuration as the relevant modes get amplified due to Floquet instabilites. \textit{Left:} 1-dimension with $\alpha_{init} = 0.99\alpha_c$, $\epsilon = 0.45$. \textit{Right:} 3-dimensions with $\alpha_{init} = 0.5 \alpha_c$ and $\epsilon = 0.2$. In both cases $\Lambda = 0.5$}
    \label{fig:gaussian}
\end{figure}

On the other hand, by closely looking at the initial (linear) evolution of the $\chi$ field we see two stages: a small bump arising from the noise and the bump growing exponentially. This can be seen clearly in Fig.~\ref{fig:gaussian}; where we note similar behaviour for the 1D and 3D systems. By initializing the $\chi$ field as a bump, we circumvent the first stage, without losing any information about the dynamical evolution of the two-field system. In what follows, this is exactly what we did in order to increase numerical efficiency.

\begin{figure}[h!]
    \centering
    \includegraphics[width = \textwidth]{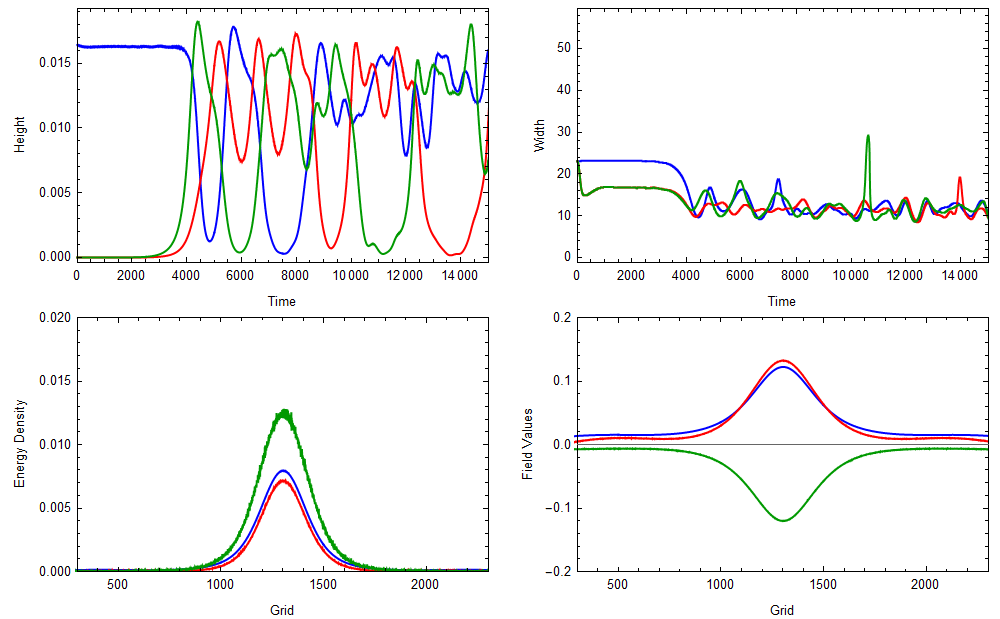}
    \caption{The decay of a single field oscillon to a three field composite oscillon. Although the presence of breathing modes doesn't allow the system to settle into a perfect symmetric state, all fields seem to oscillate around a common configuration. \textit{Top:} The evolution of the height and width of the primary (blue) $\phi$ oscillon and the two secondary oscillons (green and red)  that are formed after being initialized as a small perturbation. \textit{Bottom:} the field configurations (both in field space and energy density) at $t = 10000 (m^{-1})$ of the primary field (blue) and the two secondary fields (green and red). Here we used $\alpha = 0.99 \alpha_c$ as initialization of the primary field and $\epsilon = 0.2$.}
    \label{fig:threefields}
\end{figure}

Finally, we performed a numerical simulation for three fields. In Fig.~\ref{fig:threefields} we show the evolution of a system initialized as a single field oscillon in the primary field $\phi$, interacting with two fields initialized as  small perturbations. We again see that at late times the three fields oscillate around a common value for the width and height, forming a phase-locked three-field oscillon with internal breathing modes between the fields. In the particular example plotted we observed a phase-locked state after $t\sim 5000 (m^{-1})$ where initially one of the fields is oscillating out of phase with the other two fields. However we should report that due to the presence of the breathing modes a spontaneous change in behaviour is also possible, where a different type of phase-locking is suddenly reached. 

\subsection{Single to multi-field oscillons in  3+1D}
\label{sec:3Dfloquet}

We now move to the the analogous three-dimensional case, anticipating full lattice simulation of preheating in three dimensions. Deferring the full lattice simulation for a future publication, we instead start with a single field spherically symmetric oscillon and study the transition to the multi-field configuration, similarly to Section~\ref{sec:1Dfloquet}.

\begin{figure}
    \centering
    \includegraphics[width = \textwidth]{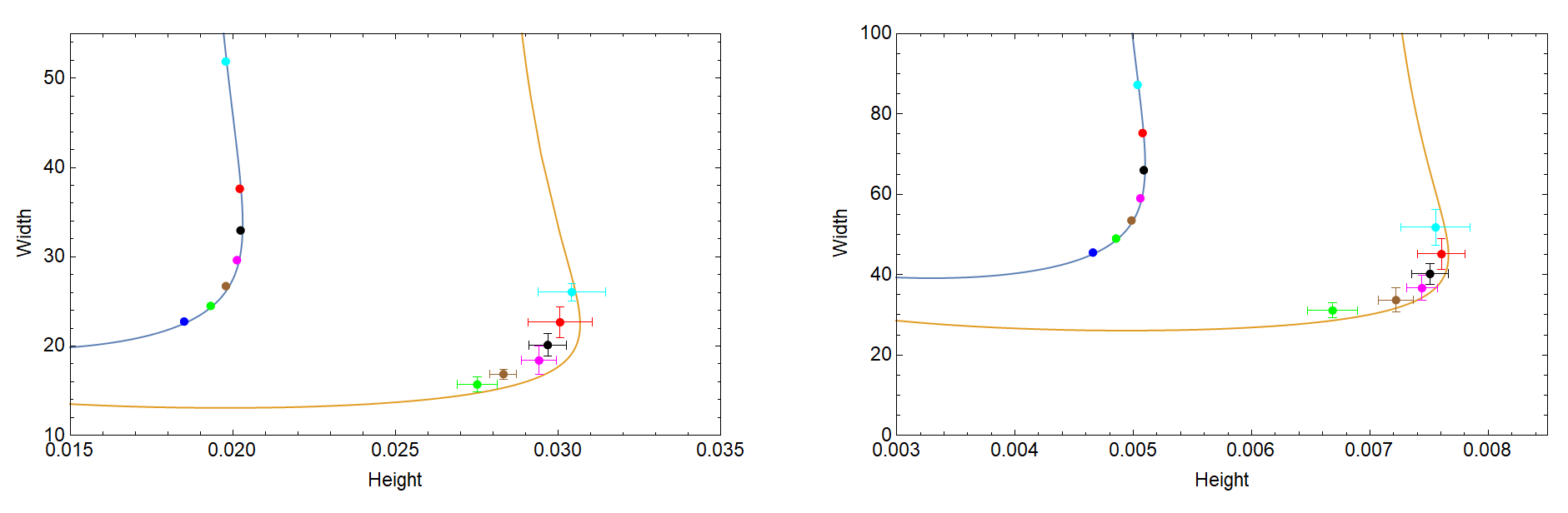}
    \caption{The evolution of the "perfect" single field oscillon, initially localized on the $\Lambda = 0$ curve (red) towards a two-field configuration falling approximately on the $\Lambda = 0.5$ curve (brown), agreeing with our theoretical predictions. For the final state, we average over the two fields and over various time slices to account for the effect of the breathing modes (hence, the error bars). The dots correspond to various initial conditions with $\alpha = 0.4, 0.45, 0.5, 0.55, 0.6, 0.65, 0.7 (\alpha_c)$ (bottom to top). \textit{Left:} $\epsilon = 0.2$. \textit{Right:}$ \epsilon = 0.1$. Note that in both cases only the single-field oscillon with $\alpha = 0.4 \alpha_c$ (blue dot) does not evolve towards a composite solution.}
    \label{fig:3Devolution}
\end{figure}

Specifically we initialize the $\phi$ field in the single oscillon configuration with $\epsilon=0.1, 0.2$ and several values of $\alpha$ and the $\chi$ fields as a small localized overdensity (bump), as described in Section~\ref{sec:1Dfloquet}. In Fig.~\ref{fig:3Devolution} we show the evolution of the width and height of the single field oscillon towards the two-field solution at the end of the simulation. Clearly, similarly to the 1-dimensional case, the oscillon moves towards a composite structure. 
As can be seen from the error bars, we see that the resulting configuration is much closer to the analytically constructed form of a two-field oscillon than its one-dimensional counterpart. In other words, the internal breathing mode exists, but is suppressed. We believe that the larger size of the internal breathing mode for one spatial dimension is connected to the increased stability for one-dimensional oscillons, compared to their three-dimensional counterparts. Otherwise, the results are qualitatively equivalent to the 1-dimensional case as represented in Fig.~\ref{fig:1devolution}. It is noteworthy that we need to initialize the oscillon with a value of $\alpha$ much closer to $\alpha_c$ in order to see decay to a two-field solution in 1D: $\alpha_i \sim 0.95 \alpha_c$ in 1D and $\alpha_i \sim 0.45 \alpha_c$ in 3D; where the subscript $i$ indicates the initial value of $\alpha$ of the background oscillon. This fact can be understood heuristically from the point of view of Floquet theory.

\subsection{Analysis of results in terms of Floquet theory}
\label{sec:floquet}
\begin{figure}
    \centering
    \includegraphics{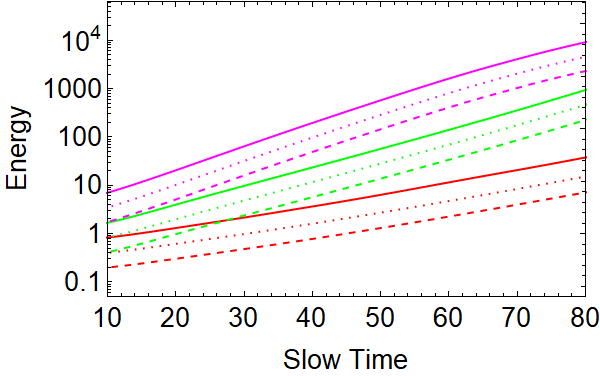}
    \caption{The amplification in the localized energy of the secondary field $\chi$ at early times in 3D on a log scale, with time rescaled as $\tau = \epsilon^2 t$ and $\alpha = 0.5 \alpha_c$ (red), $\alpha = 0.6 \alpha_c$ (green), $\alpha = 0.75 \alpha_c$ (magenta). For each value of $\alpha$ we simulated three different values of $\epsilon$: 0.05 (solid) 0.1 (dotted) and 0.2 (dashed). As expected, the amplification with respect to $\tau$ is independent of $\epsilon$ since the different linetypes are parallel to each other. There is a dependence on $\alpha$ which is allowed by Eq.~\eqref{eq:floquetexp}. Here we used $\Lambda=0.5$}
    \label{fig:energyexponential}
\end{figure}

In this subsection we explore  the emergence of the multi-field oscillons as observed in the previous sections using semi-analytic techniques. We  do this explicitly for two fields, but the intuition naturally extends to more fields. To first order in perturbations, the equation of motion of the secondary field $\chi$ is
\beq
\Ddot{\chi} -\nabla^2 \chi + \chi - \Lambda \phi^2 \chi = 0 \, ,
\label{eq:firstorderperteq}
\eeq
Since $\chi$ is a small perturbation at early times, this equation governs its dynamics. Also, to first order in $\epsilon$ we write $\phi =\epsilon \Phi_0(x) \cos(\omega t)$. Switching to Fourier space and applying the convolution theorem we obtain
\beq
\Ddot{\chi_k} + k^2 \chi_k + \chi_k - \epsilon^2 \Lambda \cos(\omega t)^2 \int \frac{d^dk}{(2\pi)^d} \Psi(k - k') \chi_{k'} = 0 \, ,
\label{eq:firstorderperteqfour}
\eeq
where $\Psi(k)$ is the Fourier transform of $[\Phi_0(x)]^2$. This equation can in principle be solved using Floquet theory, where the solutions are of the form $\chi_k(t) = e^{\mu_k t} P(t)$,  where $P(t)$ is a periodic function. However the inhomogeneity of the background oscillon couples all different k-modes through the convolution term which makes the challenge of finding the Floquet exponents impossible. To understand the basic dynamics, we investigate the constant amplitude limit  $\Phi_0(x) = \Phi_0(0)$, and incorporate the effect of the spatial extent of the oscillon heuristically. In this limit the wavenumbers decouple and it is a simple exercise to find the Floquet exponents $\mu_k$. In appendix ~\ref{app:floquetexponent} we show how to find an approximate solution for the functions $\chi_k(t)$. They are of the form $\chi_k(t) \propto Re\{\cos(\omega t) e^{i\mu_k \tau}\}$ where $\tau = \epsilon^2 t$ is slow time and

\beq
\mu_k^2 ={1\over 4} {\left (\alpha +  {k^2\over \epsilon^{2}} - \frac{3}{4}\Lambda \Phi_0^2 \right )
\left (\alpha +  {k^2\over \epsilon^{2}} - \frac{1}{4}\Lambda \Phi_0^2 \right )} \, ,
\label{eq:exponentapp}
\eeq
where $\Phi_0$ is the rescaled height of the background oscillon at the origin and $\omega = {1 - \alpha \epsilon^2/2}$ its frequency. If Eq.~\eqref{eq:exponentapp} has imaginary solutions the corresponding mode will grow exponentially. Eq.~\eqref{eq:exponentapp} makes it explicit that a two-field configuration will not emerge for repulsive interactions (negative $\Lambda$), so in what follows we  focus on attractive interactions. Note that if we find an unstable mode for some $\epsilon_1$ ($g = 1/\epsilon^2$ is the parameter multiplying the sextic term in our Lagrangian) at a wavenumber $k_1$, a different choice of $\epsilon_2$ will have an unstable mode with the same Floquet exponent at $k_2 = \frac{\epsilon_2}{\epsilon_1} k_1$. Since the oscillon width scales like $R_{osc} \propto 1/\epsilon$, the range of validity of the homogeneous approximation also changes with $\epsilon$. This explains why the decay of a single field oscillon into a two-field configuration does not seem to depend on the exact value of the small parameter\footnote{This may seem somewhat surprising since the height of the background oscillon scales with $\epsilon$. Note however that the actual timescales of decay scale with $\epsilon^2$. So although the oscillon will eventually decay to a two-field configuration, the process will take more time for smaller $\epsilon$.} $\epsilon$ (see e.g. Fig.~\ref{fig:3Devolution}); though it might depend $\alpha$ and $\Phi_0$ (these quantities are linked via the  oscillon profile equation; Eq.~\eqref{eq:prof}). This line of argument is also valid in the full inhomogeneous Floquet analysis, although it is impossible to find exact analytical expressions for the Floquet exponents in this case. The reasoning is confirmed in Fig.~\ref{fig:energyexponential} where we plot the amplification of the energy in the $\chi$ field at early times for some of the simulations we performed in three dimensions. As expected, the amplification of the field with respect to slow time $\tau$ is independent of $\epsilon$. We can use Eq.~\eqref{eq:floquetexp} to find a crude estimate of the amplification of the localized energy which goes like $E_{\chi} \propto |\chi_{k, 0}|^2 e^{2 \mu_k \tau}$, where in general we found $\mu_k \sim \mathcal{O}(0.1)$ in units of $m$, agreeing with what we observe in Fig.~\ref{fig:energyexponential}.
\par
\begin{figure}[h]
    \centering
    \includegraphics[width = \textwidth]{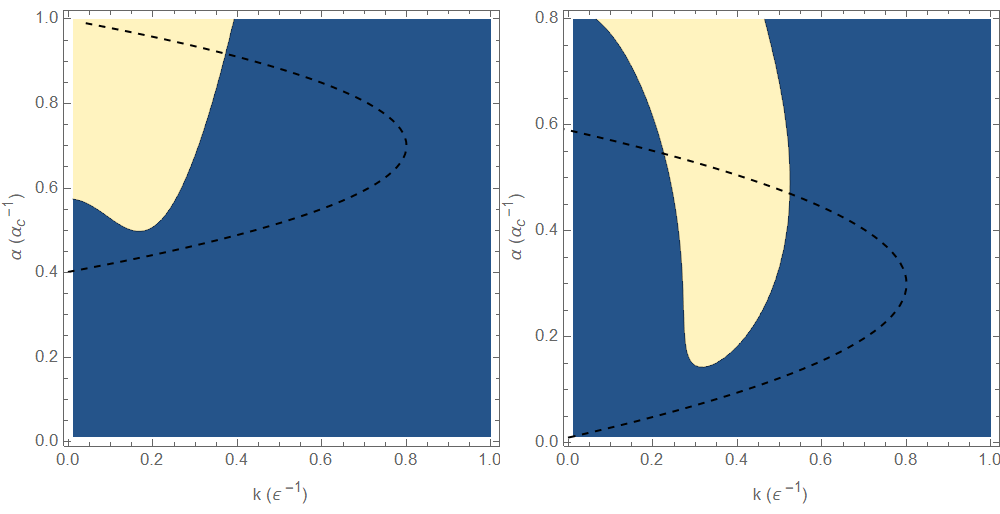}
    \caption{The instability bands for different wavenumbers $k$ and different background oscillons parametrized by $\alpha$ in the homogeneous background approximation. Here we fixed $\Lambda = 0.5$. The bands look somewhat similar in 3D (left) and 1D (right), however due to the difference in shapes of the background oscillon, this approximation is valid over a larger region of parameter space in 3D. This is shown schematically by the black dashed line: the approximation is valid for the parameter space to the right of the line, where the oscillons are wider.}
    \label{fig:floquet1dvs3d}
\end{figure}

\begin{figure}[ht!]
    \centering
    \includegraphics[width = \textwidth]{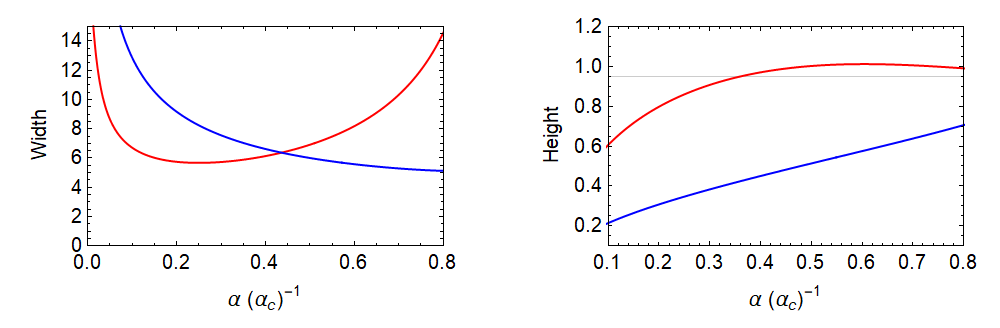}
    \caption{The dependence of the features of the background oscillon on $\alpha$ for 3D (red) and 1D (blue). We see that in three dimensions the asymptotic solution where width $\sim \infty$ and height $\sim \sqrt{9/10}$, is approximately valid for smaller $\alpha$. This explains our observations in the previous sections since Eq.~\eqref{eq:floquetexp} applies more accurately in this regime.}
    \label{fig:widthheight1d3d}
\end{figure}

 Eq.~\eqref{eq:exponentapp} is valid in both 1D and 3D. In Fig.~\ref{fig:floquet1dvs3d} we show the main instability band for different values of $\alpha$ and wavenumber $k$, with $\Lambda = 0.5$, for both three dimensions (left) and one dimension (right). 
The numerics of the previous sections have shown that in one dimension a single field oscillon settles into a two-field solution only when $\alpha \geq 0.95 \alpha_c$; while in three dimensions this happens for $\alpha \geq 0.45 \alpha_c$. This is merely a reflection of the invalidity of the approximation we have taken here. Since the oscillon has a width of order $R_{osc} \sim 1/\epsilon$, this approximation breaks down for some $k \ll \epsilon$. Correspondingly, when for a given $\alpha$ the oscillon solution is very wide, the approximation is valid for a larger range of $k-$modes. This helps us understand the difference between the 1D and 3D simulations. In Fig.~\ref{fig:widthheight1d3d} we plot the dependence of the width and height of the oscillon solution on $\alpha$. Clearly, the three-dimensional solution has a larger width than the one dimensional solution for $\alpha \sim 0.45 \alpha_c$ (also see Fig.~\ref{fig:heightwidth}). In Fig.~\ref{fig:floquet1dvs3d} this information is conveyed by the black line going through the instability band: the approximation is only valid for wavenumbers that are to the right of this line. 
Note that the line drawn here is merely schematic.

\section{Summary and Discussion}
\label{sec:Summary}

Despite the oscillons' ubiquity in non-linear scalar field theories and their assumed presence in the early Universe, very few multi-field oscillons have been found and studied. In the present work we explored a symmetric system comprised of two interacting scalar fields and showed how genuine  two-field oscillon solutions can be constructed. 

We found two-field oscillon solutions in potentials with either an attractive or a repulsive interaction term. They are qualitatively similar to ``flat-top'' oscillons found in Ref.~\cite{Amin:2010jq}, with quantitative differences that depend on the sign and strength  of the interaction term. 
The oscillons that emerge in the presence of an attractive non-linearity can be both taller (reaching higher central field values) and also narrower (having smaller width) than their single-field counterparts. 
In the case of repulsively interacting fields, the range of possible oscillon amplitudes shrinks, until the repulsive term is strong enough to completely forbid   the existence of oscillons, at least within the small amplitude two-timing framework.

The stability of non-linearity-supported localized structures can be assessed through the Vakhitov-Kolokolov (V-K) stability criterion. We formally extended the VK criterion, in order to be used in  two-field systems. We checked the VK criterion against numerical simulations (assuming spherical symmetry throughout the evolution), finding excellent agreement between semi-analytical and fully numerical results. 
Our current proof holds for symmetric potentials and the  generalization to arbitrary multi-field oscillons  is left for future work. Furthermore, we explored the basin of attraction of stable oscillons, by perturbing unstable initial configurations. We found that, depending on the size of the initial perturbation, the unstable oscillons can either completely disperse, or relax to a stable oscillon configuration with a smaller width and larger height.

Since oscillons are long-lived configurations albeit with a finite lifetime (even classically), we adapted the methods used in the literature to compute the emitted scalar radiation and thus estimate the oscillon lifetime. The results we found are qualitatively and quantitatively similar to the single-field case, showing that the longevity of multi-field oscillons can be --at least-- comparable to their single-field counterparts. We showed that the results for the two-field system can be naturally extended to models with an arbitrary number of fields, where each field has the same quadratic - quartic - sextic potential and all fields are pairwise coupled. We were able to construct multi-field oscillons, under certain conditions for the couplings, and prove their stability, by drawing formal analogies to the single-field case. This opens the way for arbitrarily dense oscillons, comprised of multiple fields.

Finally, we studied the emergence of these multi-field oscillons. We performed  simulations on an expanding lattice in one spacial dimensions, in a preheating-like scenario. We showed  that the fragmentation of the inflaton proceeds in three distinct stages: the inflaton fragments into single field oscillons; the single-field oscillons amplify fluctuations in the spectator field(s); composite oscillons emerged. Prompted by these observations we analysed the decay of single field oscillons into multi-field oscillons, both in one and three spatial dimensions, in Minkowski space. The decay can be understood semi-quantitatively using floquet theory. Since the emergence of multi-field oscillons in three spatial dimensions follows the one-dimensional behaviour in Minkowski space, we expect that multi-field oscillons will also form in  full three-dimensional simulations on an expanding lattice, which is left for a future publication.

Overall, our current work provides analytical tools for studying multi-component oscillons and opens up several avenues for future work, such as relaxing the symmetry structure of the potential and the assumption of spherical symmetry.

\acknowledgements{
We thank Ana Ach\'ucarro for useful discussions and Mustafa Amin for suggestions regarding the lifetime of oscillons.
We acknowledge support
from IFAE, which is partially funded by the CERCA program of the Generalitat de Catalunya.
EIS  acknowledges support from the Dutch Organisation for Scientific Research (NWO).
EIS also acknowledges support of a fellowship from ``la Caixa'' Foundation (ID 100010434) and from the European Union's Horizon 2020 research and innovation programme under the Marie Sk\l odowska-Curie grant agreement No 847648. The fellowship code is LCF/BQ/PI20/11760021. The research leading to these results has received funding from the ESF under the program Ayudas predoctorales of the Ministerio de Ciencia e Innovaci\'on  PRE2020-094420.

 }

\appendix
\section{The V-K criterion}
\label{app:VK}

We present the derivation of the V-K stability criterion, applied first to the case of two-field oscillons and afterwards generalized to $N$ fields.

\subsection{Two fields}
For clarity, we will separately analyze the two cases of oscillons where the two fields oscillate either in phase ($a=b$) or out of phase ($a=-b$).

\subsubsection{
Two-field system oscillating in phase
}

For oscillons comprised of the two fields oscillating in phase, $a=b$, the system of  Eqs.~\eqref{eq:linearper} becomes
\beq
\begin{split}
\partial_t^2\delta + \delta - (\partial_r^2 + \frac{2}{r}\partial_r)\delta - \epsilon^2 a^2(3 + \Lambda)\cos^2(\omega t) \delta +\epsilon^4 5ga^4\cos^4(\omega t) \delta-\epsilon^2 2\Lambda a^2 \cos(\omega t)\Delta=& 0 \, , \\
\partial_t^2\Delta + \Delta - (\partial_r^2 + \frac{2}{r}\partial_r)\Delta - \epsilon^2 a^2(3 + \Lambda)\cos^2(\omega t) \Delta +\epsilon^4 5ga^4 \cos^4(\omega t) \Delta-\epsilon^2 2\Lambda a^2 \cos(\omega t)\delta=& 0  \, .
\end{split}
\label{eq:linearper1}
\eeq
While Eqs.~\eqref{eq:linearper1} is rather complicated, the symmetries of the initial Lagrangian allow us to simplify the stability analysis, by  introducing the new variable $\xi(x, t) = \delta(x, t) + \Delta(x, t)$, which reduces Eqs.~\eqref{eq:linearper1} to 
\beq
\partial_t^2\xi + \xi - (\partial_r^2 + \frac{2}{r}\partial_r)\xi -\epsilon^2 3a^2(1 + \Lambda)cos^2(\omega t) \xi +\epsilon^4 5ga^4cos^4(\omega t) \xi = 0
\label{eq:linearper1simp}
\eeq
This equation only depends on  $\xi(x, t)$, greatly simplifying our calculations. Since we are interested in perturbations that are about the same size as the oscillon itself we perform the same change of variable as before $r \rightarrow \rho = \epsilon r$. 
Furthermore, we expect  perturbations to oscillate near the oscillon frequency. To capture the growth or decay of the perturbation we should also introduce a new ``slow" time variable $\tau = \epsilon^2 t$. We also introduce a time-scale related to the corrected oscillation frequency of the oscillon $T = \omega t$.
As before we expect behaviour on both time scales, so $\xi(x, t) \rightarrow \xi(\rho, T, \tau)$, and partial time-derivatives become full derivatives. Looking for solutions that can be written as a power series in $\epsilon\ll 1$
\beq
    \xi(\rho, T, \tau) = \sum_{n = 0}\epsilon^n \xi_{n}(\rho, T, \tau)
\label{eq:expper}
\eeq
allows for a perturbative analysis of Eq.~\ref{eq:linearper1simp}. The zeroth and first order equations are respectively
\beq
\partial_T^2 \xi_0 + \xi_0 = 0
\label{eq:0order}
\eeq
and
\beq
\partial_T^2 \xi_1 + \xi_1 = -\left(\partial_T\partial_{\tau} - \alpha \partial_T^2 - \left(\partial_{\rho}^2 + \frac{2}{\rho}\partial_{\rho}\right ) - 3 a^2 (1 + \Lambda) (\cos{T})^2 + 5 a^4 (\cos{T})^4\right)\xi_0
\, .
\label{eq:1order}
\eeq
The  general solution of Eq.~\eqref{eq:0order}  has the form
\beq
\xi_0(\rho, T, \tau) = u(\rho, \tau) \cos{T} + v(\rho, \tau) \sin(T) 
\, ,
\label{eq:sol0}
\eeq
where the  functions $u(\rho, \tau)$ and $v(\rho, \tau)$ capture the potential growth of the perturbation on timescales of order $\tau$. The V-K criterion provides a simple way to distinguish cases where perturbation   grow exponentially based on the form of the oscillon itself. Namely, inserting Eq.~\eqref{eq:sol0} into Eq.~\eqref{eq:1order} and eliminating secular terms on the right-hand side leads to  equations for $u$ and $v$
\begin{eqnarray}
\partial_{\tau}u &=& H_1 v \,  , 
\\
\partial_{\tau}v &=& - H_2 u \, ,
\end{eqnarray}
where $H_1$ and $H_2$ are Hermitian, linear operators defined as 
\begin{eqnarray}
\label{eq:A9}
H_1 =  \alpha  - (\partial_{\rho}^2 + \frac{2}{\rho}\partial_{\rho}) - \frac{3}{4}(1 + \Lambda)a^2 + \frac{5}{8} a^4 &=& 0 \, ,
\\
\label{eq:A10}
H_2 =  \alpha  - (\partial_{\rho}^2 + \frac{2}{\rho}\partial_{\rho}) - \frac{9}{4}(1 + \Lambda)a^2 + \frac{25}{8} a^4 &=& 0 \, .
\end{eqnarray}
Separating variables as $u(\rho,\tau) \to u(\rho) e^{\Omega \tau}$ and $v(\rho,\tau) \to v(\rho) e^{\Omega \tau}$
 the problem reduces to the linear equation
\beq
\Omega^2 u = - H_1 H_2 u \, .
\label{eq:eigenvalueprob}
\eeq
The question of whether the oscillon is stable to general long-wavelength perturbations is thus reduced to an eigenvalue problem. If the operator $-H_1H_2$ has at least one positive eigenvalue $\Omega^2 > 0$, perturbations can grow and the oscillon will in general be unstable. If not, perturbations simply oscillate, leading to an oscillon that is stable, within the limits of the perturbative expansion used to construct it. The problem can be solved using a similar procedure as Vakhitov and Kolokolov \cite{VK}.
We do not present the entire proof here, as its intricacies will add little to the main goal of analyzing two-field oscillons.
 The criterion for stability states that $max(\Omega^2) < 0$ if and only if $dN/d\alpha > 0$, where
 $N$ is defined in Eq.~\eqref{eq:Ndef}.
 
Before proceeding, we must make a final remark about the validity of this derivation. The criterion we presented above only states whether perturbation of the form $\xi(x, t) = \delta(x, t) + \Delta(x, t)$ will grow. In principle, the growth of the perturbation might only be present in $\delta$ or $\Delta$. However, since the system is completely symmetric under exchange of the fields, we can only conclude that, if $\xi(x, t)$ grows, both $\delta(x, t)$ and $\Delta(x, t)$ will grow exponentially. The criterion should therefore be valid for the full two-field system and can be used as an indicator for the stability of two-field oscillons, at least ones that are comprised of interchangeable fields.

\subsubsection{Two-field system oscillating out of phase}

For oscillons comprised of the two fields oscillating out of phase, $a=-b$, the system of  Eqs.~\eqref{eq:linearper} becomes
\beq
\begin{split}
\partial_t^2\delta + \delta - (\partial_r^2 + \frac{2}{r}\partial_r)\delta - \epsilon^2 a^2(3 + \Lambda)cos^2(\omega t) \delta +\epsilon^4 5ga^4cos^4(\omega t) \delta+\epsilon^2 2\Lambda a^2 cos^2(\omega t)\, \Delta= 0 \, ,
 \\
\partial_t^2\Delta + \Delta - (\partial_r^2 + \frac{2}{r}\partial_r)\Delta - \epsilon^2 a^2(3 + \Lambda)cos^2(\omega t)\Delta +\epsilon^4 5ga^4cos^4(\omega t) \Delta+\epsilon^2 2\Lambda a^2 cos^2(\omega t)\, \delta= 0 \, .
\end{split}
\label{eq:linearper2}
\eeq
Now, instead of looking for unstable modes in the combination $\xi=\delta+\Delta$, we introduce the corresponding variable $\psi \equiv \delta-\Delta$ . By now subtracting the equations in \eqref{eq:linearper2} The corresponding equation for $\psi$ becomes
\beq
\partial_t^2\psi + \psi - (\partial_r^2 + \frac{2}{r}\partial_r)\psi -\epsilon^2 3 a^2(1 + \Lambda)cos^2(\omega t) \psi +\epsilon^4 5ga^4cos^4(\omega t) \psi = 0
\, .
\label{eqlinearpersimp2}
\eeq
We see that the equation of motion for $\psi$ in the case of $a =-b$ is identical to the one for $\xi$ for $a=b$. 
The rest of the derivation follows exactly the same steps as before so we won't repeat them here. Hence the stability of the oscillon will be independent of the phase ($0$ or $\pi$) between the two fields,
resulting in exactly the same V-K criterion as before. Namely, there are unstable perturbations of the oscillon if and only if $dN/d\alpha > 0$, where $N$ is defined in Eq.~\eqref{eq:Ndef}.
\\
In this section we have derived a nontrivial extension of the V-K criterion for assessing the stability of oscillons in this coupled system to long wave-length perturbations (perturbations about the same size as the oscillon). It is, to our knowledge, the first time that the criterion has been derived for multi-component oscillons.

\begin{figure}[ht!]
    \centering
    \includegraphics[width=1\textwidth]{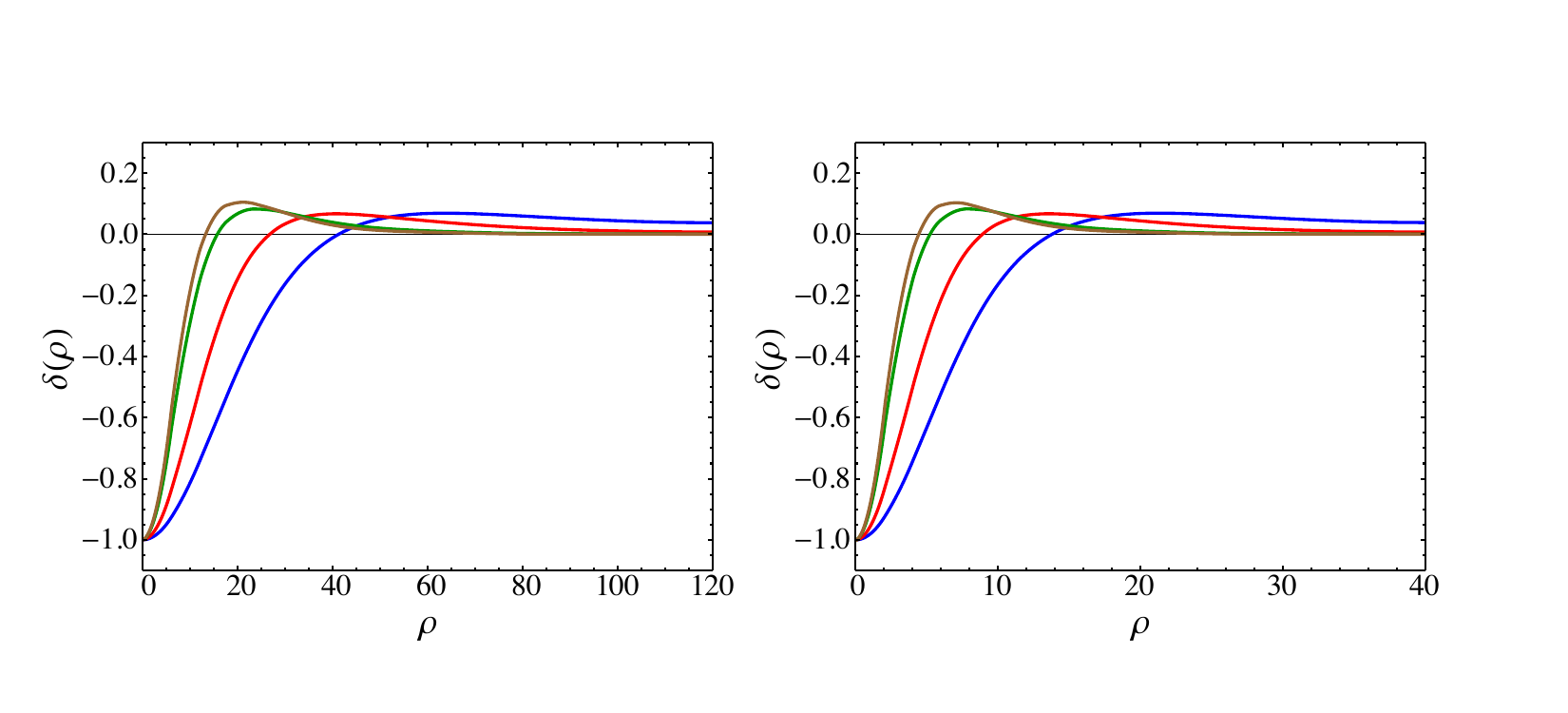}
    \caption{
    {\it Left}: The profiles of the most unstable modes of the corresponding oscillons with $\Lambda=-0.5$ and
 $\alpha= 0.00028,0.00071,  0.0025,0.0046, $ (blue, red, green and brown respectively). These are calculated numerically using the variational principle.
 {\it Right}: The profiles of the most unstable modes of the corresponding oscillons with $\Lambda=0.5$ and
 $\alpha=0.0025, 0.0064, 0.0225, 0.04 $ (blue, red, green and brown respectively).
 The profiles were rescaled such that $\delta(\rho=0)=1$. 
We see that the size of the unstable modes are similar to the size of the oscillon profile, shown in Fig.~\ref{fig:profiles}.
 }
    \label{fig:unstablemodes}
\end{figure}

Fig.~\ref{fig:unstablemodes} shows the most unstable modes computed by solving the eigenvalue problem of Eq.~\eqref{eq:eigenvalueprob} using the variational method for $\Lambda=-0.5$ and $\Lambda=0.5$, corresponding to repulsive and attractive interactions respectively. We choose some values of $\alpha$ for each case, which give $\Omega^2>0$ in Eq.~\eqref{eq:eigenvalueprob} and thus lead to (radially) unstable oscillon solutions.
 We see that the spatial size of the unstable fluctuations is indeed similar to the width of the corresponding oscillons (see e.g. Fig.~\ref{fig:profiles}).

\subsection{$N$ fields}
By solving the profile equations of Eq.~\eqref{eq:profilesN} we have implicitly found oscillon solutions of the form $\phi_{i, osc} = \epsilon a_i \cos(\omega t)$, where $\omega = 1 - \alpha {\epsilon^2}/{2}$. To assess the stability of these solutions we add a small perturbation $\delta_i$ to the oscillon solution; $\phi_i = \phi_{i, osc} + \delta_i$. Plugging this ansatz into  Eq.~\eqref{eq:motionN}, linearizing and ignoring source terms
\beq
\begin{split}
\partial_t^2 \delta_ i - (\partial_r^2 + \frac{2}{r}\partial_r)  \delta_i + \delta_i =\left( 3 \phi_{i, osc}^2 - g \phi_{i, osc}^4 + 
\Lambda \sum_{j \neq i}\left( \phi_{j, osc}^2 + 2 \phi_{i, osc} \phi_{j, osc}\frac{\delta_j}{\delta_i}\right)\right)\delta_i
\end{split}
\label{eq:perturbeqN}
\eeq
Now, assuming that $a(\rho)$ is a solution of Eq.~\eqref{eq:profilessimpleN}, we can set $n<N$ of the $a_i = a$ and $(N-n)$ of the $a_i = -a$. This configuration obviously solves the system of profile equations in Eqs.~\eqref{eq:profilesN}. If $a_i = a$ the equation for $\delta_i$ becomes (omitting all  factors of $\epsilon$ for clarity)
\beq
\begin{split}
\partial_t^2 \delta_ i - (\partial_r^2 + \frac{2}{r}\partial_r)  \delta_i + \delta_i =&\left( 3 a^2\cos^2(\omega t) - g a^4\cos^4(\omega t) + \Lambda (N-1)a^2\cos^2(\omega t)\right)\delta_i 
\\
&+2 \Lambda a^2\cos^2(\omega t) \sum_{j \neq i} \delta_j - 2 \Lambda a^2\cos^2(\omega t) \sum \Delta_j
\, ,
\end{split}
\label{eq:perturbeqNp}
\eeq
while for $a_i = -a$ the same equation becomes
\beq
\begin{split}
\partial_t^2 \Delta_ i - (\partial_r^2 + \frac{2}{r}\partial_r)  \Delta_i + \Delta_i  =& \left( 3 a^2\cos^2(\omega t) - g a^4\cos^4(\omega t) + \Lambda (N-1)a^2\cos^2(\omega t)\right)\Delta_i 
\\&+ 2 \Lambda a^2\cos^2(\omega t) \sum_{j \neq i} \Delta_j - 2 \Lambda a^2\cos^2(\omega t) \sum \delta_j
\, ,
\end{split}
\label{eq:perturbeqNm}
\eeq
where we've renamed the perturbations $\delta_i \rightarrow \Delta_i$ if $a_i = -a$ for clarity. Adding the $n$ equations of the form \eqref{eq:perturbeqNp} and the $N - n$ equations of the form \eqref{eq:perturbeqNm} together respectively we obtain the equations
\beq
\begin{split}
\partial_t^2 \Omega - (\partial_r^2 + \frac{2}{r}\partial_r) \Omega + \Omega =&
\left( 3 a^2\cos^2(\omega t) - g a^4\cos^4(\omega t) + \Lambda (N-1)a^2\cos^2(\omega t)\right)\Omega 
\\
&+ 2 \Lambda (n - 1)a^2\cos^2(\omega t) \Omega - 2 \Lambda n a^2\cos^2(\omega t) K
\label{eq:perturbeqNp2}
\end{split}
\eeq
and
\beq
\begin{split}
\partial_t^2 K- (\partial_r^2 + \frac{2}{r}\partial_r)  K+ K =&
\left( 3 a^2\cos^2(\omega t) - g a^4\cos^4(\omega t) + \Lambda(N-1)a^2\cos^2(\omega t)\right)K \\&+ 2 \Lambda((N - n - 1)a^2\cos^2(\omega t) K - 2 \Lambda(N - n) a^2\cos^2(\omega t) \Omega
\, ,
\label{eq:perturbeqNm2}
\end{split}
\eeq
where we've introduced variables $\Omega = \sum \delta_i$ and $K = \sum \Delta_i$. Finally, we substract Eq.~\eqref{eq:perturbeqNm2} from Eq.~\eqref{eq:perturbeqNp2} and define the variable $\Xi = \Omega - K$ to obtain
\beq
\partial_t^2 \Xi - (\partial_ r^2 + \frac{2}{r}\partial_r) \Xi + \Xi = 
\left( 3(1 +\Lambda(N-1))  a^2\cos^2(\omega t) - g a^4\cos^4(\omega t)\right)\Xi
\eeq
From here the derivation of the instability bands of $\Xi$ is analogous to the derivation performed in section \ref{sec:VK}. The difference being that $\Lambda \rightarrow \Lambda(N-1)$ with respect to Eq.~\eqref{eq:linearper1simp}. However, this is exactly the change that occurs in the effective profile equation \eqref{eq:profilessimpleN}. The V-K criterion therefore still applies and the multi-field oscillon is stable if and only if $dN/d\alpha > 0$, where
\beq
N = \int a^2(\rho) d^3\rho \, ,
\eeq
which is identical to Eq.~\eqref{eq:Ndef}.

\section{Aprroximate Floquet exponents}
\label{app:floquetexponent}
We provide an approximate analytic method for computing the Floquet exponents of the equation
\beq
\Ddot{\chi_k} + (k^2 + 1)\chi_k - \Lambda \epsilon^2 \Phi_0^2 \cos^2(\omega t) \chi_k = 0 \, .
\label{eq:homopert}
\eeq
Which is just the limit of Eq.~\eqref{eq:firstorderperteqfour} in which the oscillon is approximated as a homogeneous background. We can find perturbative solutions by introducing two time scales: the period $T = \omega t = {1 - \alpha \epsilon^2/2} t$ and the slow time $\tau = \epsilon^2 t$. We also assume that the solution has characteristic behaviour on both timescales, similar to the two-timing analysis, meaning
 $\chi_k(t) = \chi_k(T, \tau)$. Finally, we are interested in modes for which $k = \kappa \epsilon$ where $\kappa =  \mathcal{O}(1)$. Eq.~\eqref{eq:homopert} becomes
\beq
\left (1 - \alpha \epsilon^2 \right ) \Ddot{\chi}_k + 2 \epsilon^2 \dot{\chi'}_k+ \left (\epsilon^2\kappa^2 + 1 \right )\chi_k - \Lambda \epsilon^2 \Phi_0^2 \cos^2(\omega t) \chi_k = 0
\eeq
where a dot denotes derivation with respect to $T$ and a prime denotes derivation with respect to $\tau$. We look for a perturbative solution $\chi_k = \chi^{(0)}_k + \epsilon \chi^{(1)}_k + \epsilon^2 \chi^{(2)}_k+ ...$\\
The $\mathcal{O}(\epsilon^0)$ gives the solution of $\chi^{(0)}_k$
\beq
\chi^{(0)}_k = A(\tau) \cos(T) + B(\tau) \sin(T)
\label{eq:chik0sol}
\eeq
The $\mathcal{O}(\epsilon^1)$ equation does not provide any additional information. The $\mathcal{O}\left (\epsilon^2\right )$ equation is
\beq
\Ddot\chi^{(2)}_k + \chi^{(2)}_k = \alpha \Ddot\chi^{(0)}_k - 2 \dot\chi'^{(2)}_k - \kappa^2 \chi^{(0)}_k + \Lambda \Phi_0^2 \cos^2(T)
\label{eq:secorderpert}
\eeq
By plugging in Eq.~\eqref{eq:chik0sol} into the right hand side of Eq.~\eqref{eq:secorderpert} and requiring that the 0th-order perturbation doesn't induce resonances in the 2nd-order perturbation we obtain the following constraints on $A(\tau)$ and $B(\tau)$

\beq
\begin{split}
    2 B' &= -\left(\alpha + k^2 - \frac{3}{4}\Lambda \Phi_0^2\right) A\\
    2 A' &= \left(\alpha +k^2 - \frac{1}{4}\Lambda \Phi_0^2\right) B\\
\end{split}
\eeq
This system of equations can be trivially solved
by applying $\partial_\tau$ on one of the two equations and using the other to elliminate the first derivative term, leading to  
\beq
A''+ \left(\alpha + k^2 - \frac{3}{4}\Lambda \Phi_0^2\right) \left(\alpha +k^2 - \frac{1}{4}\Lambda \Phi_0^2\right)A=0
\eeq
with $B$ obeying the same equation. The solutions are $A(\tau), B(\tau) \sim \cos(\mu_k \tau), \sin(\mu_k \tau)$. When $\mu_k$ is imaginary, the trigonometric functions become hyperbolic trigonometric functions, which for late times grow as $e^{Im[\mu_k]\tau}$. We thus arrive at the result expected by Floquet theory, where the Floquet exponents are given by
\beq
\mu_k^2 = \frac{(\alpha + \kappa^2 - \frac{3}{4}\Lambda \Phi_0^2)(\alpha + \kappa^2 - \frac{\Lambda}{4}\Phi_0^2)}{4} \, .
\label{eq:floquetexp}
\eeq
\begin{figure}[h!]
    \centering
    \includegraphics[width = \textwidth]{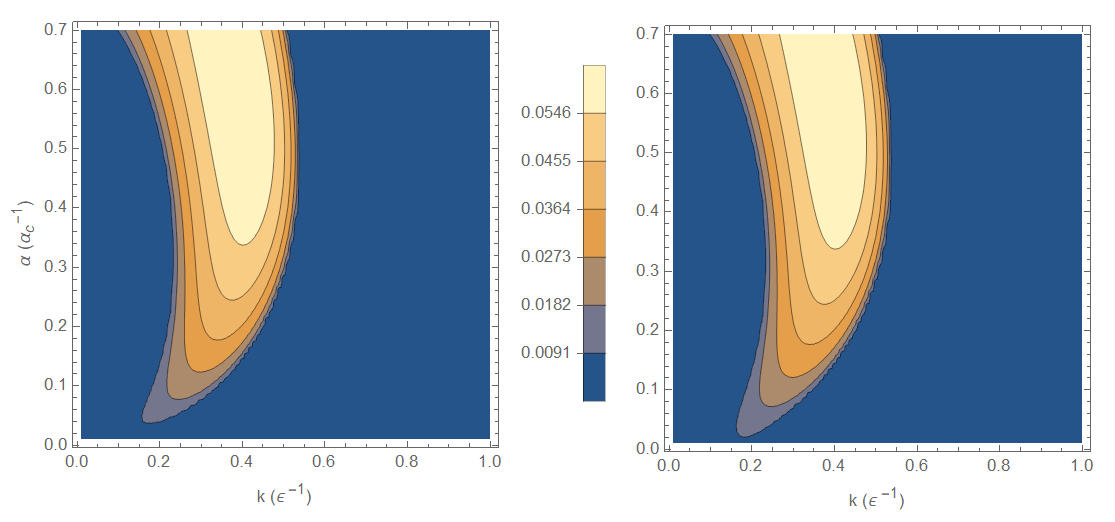}
    \caption{Comparison of the Floquet instability bands within the homogeneous approximation, computed with a full numerical analysis (left) for $\epsilon = 0.1$ and computed using Eq.~\eqref{eq:floquetexp} (right). The instability bands agree extremely well, speaking to the validity of our analysis. Here we use $\Lambda = 0.5$.}
    \label{fig:floquetcomparison}
\end{figure}
This is the expression given in the main body of text. 
In Fig.~\ref{fig:floquetcomparison} we compare the instability bands computed in this approximation with the full numerical Floquet analysis of Eq.~\eqref{eq:homopert} for the case of $\epsilon=0.1$ in three dimensions. As one can see, the perturbative method highlighted in this section agrees extremely well with the full computation.

\bibliography{Composite_oscillons}

\end{document}